\documentclass[12pt,draftcls,peerreviewca,onecolumn,a4paper,dvips]{IEEEtran}
\usepackage{pifont}
\usepackage{amsfonts}
\usepackage{mathrsfs}
\usepackage{amssymb}
\usepackage{graphicx}
\usepackage{psfrag}
\usepackage{amsmath}
\usepackage{array}
\usepackage{cases}
\begin{document}

\author{\IEEEauthorblockN{Liangzhong Ruan,   Vincent K.N. Lau}
\IEEEauthorblockA{Dept. of  Electronic and Computer Engineering\\
the Hong Kong University of Science and Technology
\\Email: $\{$stevenr,
eeknlau$\}$@ust.hk}}
\title{Dynamic Interference Mitigation for Generalized Partially Connected Quasi-static MIMO Interference Channel}

\newtheorem{Thm}{Theorem}[section]
\newtheorem{Lem}{Lemma}
\newtheorem{Asm}{Assumption}[section]
\newtheorem{Def}{Definition}[section]
\newtheorem{Remark}{Remark}[section]
\newtheorem{Prob}{Problem}
\newtheorem{Alg}{Stage}
\newtheorem{Example}{Example}
\newtheorem{Cor}{Corollary}[section]

\maketitle

\begin{abstract}
Recent works on MIMO interference channels have shown that
interference alignment can significantly increase the achievable
degrees of freedom (DoF) of the network. However, most of these
works have assumed a fully connected interference graph. In this
paper, we investigate how the partial connectivity can be exploited
to enhance system performance in MIMO interference networks. We
propose a novel interference mitigation scheme which introduces
constraints for the signal subspaces of the precoders and
decorrelators to mitigate ``many" interference nulling constraints
at a cost of ``little" freedoms in precoder and decorrelator design
so as to extend the feasibility region of the interference alignment
scheme. Our analysis shows that
 the proposed algorithm can significantly increase system DoF
in symmetric partially connected MIMO interference networks. We also
compare the performance of  the proposed scheme with various
baselines and show via simulations that the proposed algorithms
could achieve significant gain in the system performance of randomly
connected interference networks.
\end{abstract}

\section{Introduction}\label{sec:intro}
Recently, there is an intense research interest in the area of
interference channels and the interference mitigation techniques.
 {\em
Interference alignment} (IA) was proposed in
\cite{firstIA1,firstIA2} to reduce the effect of multi-user
interference and is extended to deal with interference in MIMO
X-channels \cite{JafarX1} and $K$ pairs interference channels
\cite{JafarK1}. The key idea of IA is to reduce the dimension of the
aggregated interference by aligning interference from different
transmitters into a lower dimension subspace at each receiver. Using
infinite dimension extension on the time dimension (time selective
fading), it is shown that the IA could achieve the optimal
Degrees-of-Freedom (DoF) of $\frac{KN}{2}$ in
$K$-pair MIMO ergodic interference channels
\cite{JafarK1} with $N$ antennas at each node. In \cite{JafarE}, the
authors proposed the concept of {\em ergodic alignment}, which also
utilizes symbol extension exploiting time-selective fading of
interference channels.

One important challenge of IA scheme is the feasibility condition.
For instance, the IA schemes in \cite{JafarK1} requires
$\mathcal{O}((KN)^{2K^2N^2})$ dimensions of signal space to achieve
the $\frac{KN}{2}$ total DoF. To avoid such huge dimensions of
signal space, some researchers have studied IA designs for
quasi-static MIMO interference channels. With limited signal space
dimensions, the achievable DoF of each transmitter-receiver pair in
MIMO interference channels is upper bounded by $\frac{N_t+N_r}{K+1}$
(where $K$ is the number of transmitter-receiver pairs, $N_t$, $N_r$
are the number of antennas at each transmitter and receiver,
respectively) \cite{JafarDf}. Unlike the time-selective or
frequency-selective MIMO interference channels, total DoF of
quasi-static MIMO interference channel does not scale with $K$.
Furthermore, it is quite challenging to design precoders and
decorrelators that satisfy the IA requirements in limited dimension
MIMO interference channels due to the feasibility problems
\cite{JafarDf}. In \cite{JafarD1}, an iterative precoders and
decorrelators design based on alternating optimization
 is proposed for quasi-static MIMO interference channels.
 In \cite{MIMOC1,MIMOC2}, some constructive methods to design
precoders and decorrelators are proposed, but these schemes can only
achieve 1 DoF per transmitter.

In fact, the technical challenge on the feasibility issue in limited
dimension MIMO interference channels is highly related to the full
connectivity in the interference graph. In practice, the
interference channels are usually {\em partially connected} due to
path loss, shadowing as well as spatial correlation. Most of the
existing literatures have assumed fully connected MIMO interference
channels such as equal path loss and spatially uncorrelated MIMO
channels. Intuitively, partial connectivity may contribute to
limiting the aggregate interference and this may translate into DoF
gains in the system. In this paper, we are interested to study the
potential benefit of partially connectivity in a
$K$-pair MIMO interference network with quasi static
fading. There are several important technical challenges involved.

\begin{itemize}
\item {\bf How to exploit partial connections
in interference mitigation?} Traditionally, it is well-known that
partial connection (due to path loss, shadowing or spatial
correlation) is detrimental to  point-to-point MIMO performance
\cite{SDMA_0,AngleC2} because it reduces the number of spatial
channels in the MIMO link. However, in MIMO interference channels,
partial connection may also reduce the dimension of the undesired
signals (the interference), leading to possible performance
improvement. In other words, we can potentially design precoders to
exploit the partial connection property and reduce the interference
dimensions to other users.
\item {\bf Achievable DoF for partially connected MIMO interference channels.}  In order to obtain
insights on the potential benefits or degradations of partial
connections in MIMO interference channels, one would be interested
in deriving DoF bounds. Existing DoF analysis of interference
channels \cite{JafarK1,JafarDf} assumed i.i.d. MIMO fading channels
(fully connected MIMO interference channels) and it is interesting
to find out how the partial connection parameters such as node
density and spatial correlation level affect total DoF of the
system.
\end{itemize}

In this paper, we propose a novel two-stage  dynamic interference
mitigation  scheme to exploit the potential benefit of general
partial connections in limited dimension MIMO interference channels
so as to improve the network total DoF. The proposed dynamic
interference mitigation solution has two stages. The first stage
determines the stream assignment and the subspace constraints for
the precoders and the decorrelators based on the partially connected
topology such as the path loss, shadowing and spatial correlation.
The second stage determines the precoders and the decorrelators
(based on the instantaneous channel state information) over the
subspaces obtained from the first stage. Based on the proposed
dynamic interference mitigation scheme, we shall derive an
achievable DoF bound of a symmetric interference network and show
that the DoF in partially connected MIMO interference channels can
exceed the well-known DoF results of $\frac{N_t+N_r}{K+1}$ for
i.i.d. MIMO interference channels. Furthermore, we shall discuss how
the DoF gain is affected by the partial connectivity (path loss and
spatial correlation) in the system. Finally, we shall compare the
performance of the proposed scheme with various conventional
baselines, and it can be observed that the proposed scheme offers
significant performance gain over a wide range of system operating
regimes.

\section{System Model}
\subsection{General $K$-pair Partially Connected Quasi-static MIMO Interference Channels}
\label{sec:model} We consider a MIMO system with $K$ transmitter
(Tx) and receiver (Rx) pairs. Each transmitter and each receiver has
$N_t$ and $N_r$ antennas, respectively. Denote the channel fading
coefficients  from the Tx $n$ to the Rx
$m$ as $\mathbf{H}_{mn}\in\mathbb{C}^{N_r\times
N_t}$. Let $d_m$ ($\le \min(N_t,N_r)$) be the number
of data streams (DoF) transmitted by Tx-Rx  pair $m$.
The received signal $\mathbf{y}_{m}\in \mathbb{C}^{d_m}$ at Rx
$m$ is given by:
\begin{eqnarray}
\mathbf{y}_{m}=\mathbf{U}_{m}\left(\mathbf{H}_{mm}
\mathbf{V}_{m}\mathbf{x}_{m} + \sum_{n\neq m\in\{1,2,...,K\}}
\mathbf{H}_{mn} \mathbf{V}_{n}\mathbf{x}_{n}+\mathbf{z}
\right)\label{eqn:signal_1}
\end{eqnarray}
where $\mathbf{x}_m \in \mathbb{C}^{d_m}$  is the encoded
information symbol for Rx $m$, $\mathbf{U}_{m}\in
\mathbb{C}^{d_m \times N_r}$  is the decorrelator of Rx
$m$, and $\mathbf{V}_{m}\in \mathbb{C}^{N_t \times
d_m}$  is the transmit precoding matrix at the Tx
$m$. $\mathbf{z}\in \mathbb{C}^{N_r\times 1}$ is the
white Gaussian noise with unit variance. The transmit power at the
Tx $n$ is
$\mathbb{E}(||\mathbf{V}_{n}\mathbf{x}_{n}||^2)=P_n$. The channel
connectivity of the $K$-pair interference channels
$\{\mathbf{H}_{mn}\}$ is specified by the following model.
\begin{Asm}[General Partially Connected Model]\label{asm:channel}
The elements of the channel states matrices
 $\{\mathbf{H}_{mn}\in \mathbb{C}^{N_r\times N_t}\}$,
$n,m\in\{1,2,...,K\}$ are random variables following certain
 distribution and have the following properties:
\begin{itemize}
\item{\bf Independence:} {Random matrices} $\{\mathbf{H}_{mn}\}$
$n,m\in\{1,2,...,K\}$ are mutually independent.
\item{\bf Partial Connectivity at the Transmitter Side:} Define the Tx side partial connectivity as the null space of
$\mathbf{H}_{mn}$, e.g: $\mathcal{N}(\mathbf{H}_{mn})=\{
\mathbf{v}\in\mathbb{C}^{N_t\times 1}:
\mathbf{H}_{mn}\mathbf{v}=\mathbf{0}\}$.
\item{\bf Partial Connectivity at the Receiver Side:} Define the Rx side partial connectivity as the ``transposed"
null space of $\mathbf{H}_{mn}$, e.g:
$\mathcal{N}^H(\mathbf{H}_{mn})=\{ \mathbf{u}\in\mathbb{C}^{1\times
N_r}: \mathbf{H}^H_{mn}\mathbf{u}^H=\mathbf{0}\}=\{
\mathbf{u}\in\mathbb{C}^{1\times N_r}:
\mathbf{u}\mathbf{H}_{mn}=\mathbf{0}\}$. ~\hfill \IEEEQED
\end{itemize}
\end{Asm}

As a result, $\{\mathcal{N}(\mathbf{H}_{mn})\}$ and
$\{\mathcal{N}^H(\mathbf{H}_{mn})\}$ $m,n\in\{1,2,...,K\}$ are the
connection topological parameters of the general partially connected
model. This is a general model as no specific structure is imposed
on $\{\mathcal{N}(\mathbf{H}_{mn})\}$ and
$\{\mathcal{N}^H(\mathbf{H}_{mn})\}$. To help readers get some
concrete understanding on the physical scenarios when we have
partial connectivity, we shall illustrate some examples of partially
connected interference channels in the next section. Note that the
partial connectivity model imposed in Assumption~\ref{asm:channel}
contains, but is not limited to, these examples.

\subsection{Example Scenarios of Partially Connected Systems}\label{sec:sys_example}
\subsubsection{\bf Path Loss and Shadowing}
\label{sec:path} In practice, different Txs may contribute
differently to the aggregate interference  due to the heterogeneous
path loss and shadowing effects. For example, in a
$K$-pair MIMO interference network with
$N_t=N_r$, suppose some Txs and Rxs are far away from each other
when the difference between their node indices $|n-m|>L$, the path
loss and shadowing from Tx $n$ to Rx
$m$,
 is 60 dB higher than that of the direct link (from Tx $n$
to Rx $n$, $n\in\{1,2,...,K\}$) and hence, effectively, we have
$\mathbf{H}_{mn} = \mathbf{0}$, $\forall |n-m|>L$. This corresponds
to a partially connected MIMO interference channel (induced by path
loss and shadowing effects) with the {\em connection topology} given
by
$\mathcal{N}(\mathbf{H}_{mn})=\left\{\begin{array}{l}\mathbb{C}^{N_t\times 1}\mbox{ if }|n-m|>L\\
\{0\} \mbox{ otherwise}\end{array}\right.$,
$\mathcal{N}^H(\mathbf{H}_{mn})=\left\{\begin{array}{l}\mathbb{C}^{1\times
N_r}\mbox{ if }|n-m|>L\\
\{0\} \mbox{ otherwise}\end{array}\right.$.

\subsubsection{\bf Unequal Transmit and Receive
Antennas}\label{sec:unequal} When $N_t \neq N_r$, there is a rank
$|N_t-N_r|$ null space on the side with more antennas. For example,
when $N_t=4$,
$N_r=2$, denote $\mathbf{H}_{mn}=\left[\begin{array}{c}\mathbf{h}_{mn}(1)\\
\mathbf{h}_{mn}(2)\end{array}\right]$, where $\mathbf{h}_{mn}(p)$
are $1\times 4$ vectors. Hence, this corresponds to a partially
connected MIMO interference channel (induced by non-square fading
matrices) with the connection topology given by:
$\mathcal{N}(\mathbf{H}_{mn})=
(\mbox{span}(\mathbf{h}^T_{mn}(1),\mathbf{h}^T_{mn}(2)))^{\bot}$,
$\mathcal{N}^H(\mathbf{H}_{mn})=\{0\}$, $\forall
m,n\in\{1,2,...,K\}$.

\subsubsection{\bf Spatial Correlation}
\label{sec:Saptial} As shown in \cite{Local2},\cite{Local3}, in
practice, local scattering effect causes significant spatial
correlation in MISO channels. To explore the similar effect in MIMO
channels and get a direct association between the MIMO fading
channel correlation and the physical scattering environment, we
shall introduce the {\em virtual angular domain} representation for
MIMO channels
 \cite{AngleC2}.
Specifically, the MIMO fading channels in the {\em antenna domain}
$\mathbf{H}_{mn}^{ant}$ and in the {\em angular domain}
$\mathbf{H}_{mn}^{ang}$ have a one-one correspondence given by:
\begin{eqnarray}
\mathbf{H}^{ant}_{mn}=\mathbf{A}_R\mathbf{H}^{ang}_{mn}\mathbf{A}^H_T
\label{eqn:ant_vir}
\end{eqnarray}
\begin{eqnarray}
\mbox{where:
}\;\;\;\mathbf{A}_T&=&[\mathbf{e}_{N_t}(0),\mathbf{e}_{N_t}\left(\frac{1}{N_t}\right)
...,\mathbf{e}_{N_t}\left(\frac{N_t-1}{N_t}\right)] \label{eqn:at}
\\\mathbf{A}_R&=&[\mathbf{e}_{N_r}(0),\mathbf{e}_{N_r}\left(\frac{1}{N_r}\right)
...,\mathbf{e}_{N_r}\left(\frac{N_r-1}{N_r}\right)] \label{eqn:ar}
\\\mathbf{e}_N(\omega)&=&\frac{1}{\sqrt{N}}[1,e^{-j2\pi(\omega)},
e^{-j2\pi(2\omega)}...e^{-j2\pi((N-1)\omega)}]^T\label{eqn:e}
\end{eqnarray}

\begin{figure} \centering
\includegraphics[scale=0.6]{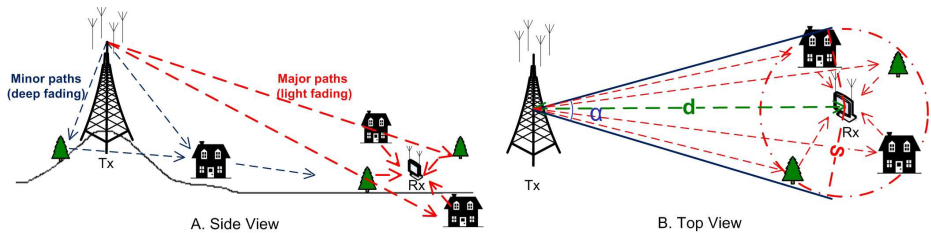}
\caption {Illustration of \emph{local scattering} effect: The
asymmetric propagation environment leads to spatial channel
correlation.} \label{fig_channel1}
\end{figure}

Given a local scattering environment with the parameter effective
scattering radius $S$ as illustrated in Fig.~\ref{fig_channel1}, the
MIMO
 fading matrix
$\mathbf{H}^{ang}_{mn}=\{h^{ang}_{mn}(p,q)\}$ $p\in\{1,2,...,N_r\}$,
$q\in\{1,2,...,N_t\}$ in the angular domain has the following
property: $h^{ang}_{mn}(p,q)=0$ with probability 1 if and only if
(Please refer to Appendix~\ref{channel})
\begin{eqnarray}
\left|\frac{\sin\theta}{2}-\frac{q}{N_t}\right|\mbox{ mod } 1>
\frac{1}{N_t},\;\forall \theta\in [\theta_{mn}-F_a(S,d_{mn}),
\theta_{mn}+F_a(S,d_{mn})]
   \label{eqn:virtual_1}
\end{eqnarray}
\begin{eqnarray}\nonumber\mbox{where }F_a(S,d_{mn})=\left\{\begin{array}{l}
 \arcsin\frac{S}{d_{mn}} \mbox{ when: }S\le d_{mn}
 \\\pi \;\;\;\; \;\;\;\; \;\;\;\; \;\mbox{ when: }S> d_{mn}\end{array}
\right.,
\end{eqnarray}
$\theta_{mn}\in(-\pi,\pi]$ is the direction from Tx
$n$ to Rx $m$, and $d_{mn}$ is the
distance between the two nodes. This spatially correlated MIMO model
is a special case of  the general partially connected MIMO
interference model. For example, in Fig.~\ref{fig_channel1}, suppose
$N_t=N_r=4$ and denote
$\mathbf{H}_{mn}^{ang}=[\mathbf{h}_{mn}(1),\mathbf{h}_{mn}(2),\mathbf{h}_{mn}(3),\mathbf{h}_{mn}(4)]$.
Suppose due to spatial correlation,
$\mathbf{h}_{mn}(1)=\mathbf{h}_{mn}(4)=\mathbf{0}$  and
$\mathbf{h}_{mn}(2),\mathbf{h}_{mn}(3)$ are randomly generated
$\mathbb{C}^{4\times1}$ vectors. $\mathcal{N}(\mathbf{H}_{mn}) =
\mbox{span}\left(\mathbf{A}^H_T\left[\begin{array}{c}1\\0\\0\\0\end{array}\right],
\mathbf{A}^H_T\left[\begin{array}{c}0\\0\\0\\1\end{array}\right]\right)$
and $\mathcal{N}^H(\mathbf{H}_{mn})=
\left(\mbox{span}\left((\mathbf{A}_R\mathbf{h}_{mn}(2))^T,
(\mathbf{A}_R\mathbf{h}_{mn}(3))^T\right)\right)^{\bot}$, $\forall
n,m\in\{1,2,...,K\}$.

\section{Algorithm Description}
In this section, we shall propose a novel dynamic interference
mitigation scheme to exploit the topological advantage due to
partial connectivity. The proposed scheme is also backward
compatible with existing IA designs when the topology is fully
connected. The algorithm dynamically determines the
data stream assignment $\mathbb{D}= \{d_1,d_2,...,d_K\}$,
$d_n\in\{0,1,...,d^{\max}_k\}$ and the associated precoders
$\mathbf{V}_n\in\mathbb{C}^{N_t\times d_n}$ and decorrelators
$\mathbf{U}_n\in\mathbb{C}^{d_n}\times N_r$, where
$d^{\max}_n$ and $d_n$ are the number of the data streams claimed by
and assigned to Tx-Rx pair $n$, respectively, $n\in\{1,2,...,K\}$,
such that:
\begin{eqnarray}
\mbox{rank}(\mathbf{U}_{n}\mathbf{H}_{nn} \mathbf{V}_{n})&=&{d_n}
\label{eqn:rank_0}
\\ \mathbf{U}_{m}\mathbf{H}_{mn} \mathbf{V}_{n}&=&\mathbf{0}, \forall
n\neq m\in\{1,2,...,K\}\label{eqn:cs_cross_0}
\end{eqnarray}

Most of the  existing works on IA have assumed a fully connected
interference topology, e.g. $\{\mathbf{H}_{mn}\}$ are all full rank.
However, as we have illustrated in Section~\ref{sec:sys_example},
 MIMO interference channels are usually partially connected due
to various physical reasons. As far as we are aware, no existing
schemes can be extended easily to exploit the potential benefit of
partially connectivity in MIMO interference
networks.
\subsection{Motivations: A Dynamic Interference Mitigation
Scheme for General Partially Connected MIMO Interference Channel}
\subsubsection{\bf The potential benefit of partial connectivity}
\label{sec:alg_benifit} We shall first illustrate the potential
benefit of partial connectivity by a simple example.

Consider a $2\times 2$, 5-pair interference network. Each Tx-Rx pair
attempts to transmit 1 data stream (i.e. $d^{\max}_n=1$, $\forall
n\in\{1,2,3,4,5\}$). If the network is fully connected (all channel
matrices are rank 2), the freedoms in each precoder
$\mathbf{v}_n\in\mathbb{C}^{2\times 1}$ and decorrelator
$\mathbf{u}_m\in\mathbb{C}^{1\times 2}$ is given by
$\mbox{dim}(\mathcal{G}(1,2))=1$, where the Grassmannian
\cite{Grass1,Grass2} $\mathcal{G}(x,y)$ denotes the set of all
$x$-dimensional subspaces in $\mathbb{C}^{y}$, and the number of
constraints induced by each cross link
($\mathbf{u}_{m}\mathbf{H}_{mn} \mathbf{v}_{n}=0$, $n\neq m$) is 1.
If we assign data streams to $k$ Tx-Rx pairs, there are in total
$2k$ freedoms in the precoders and the decorrelators and $k(k-1)$
interference alignment constraints. Hence, from the IA feasibility
condition \cite{JafarDf}, we have $k(k-1) \le 2k\Rightarrow k\le 3$.
In other words, the achievable DoF is upper bounded by 3.

\begin{figure} \centering
\includegraphics[scale=0.4]{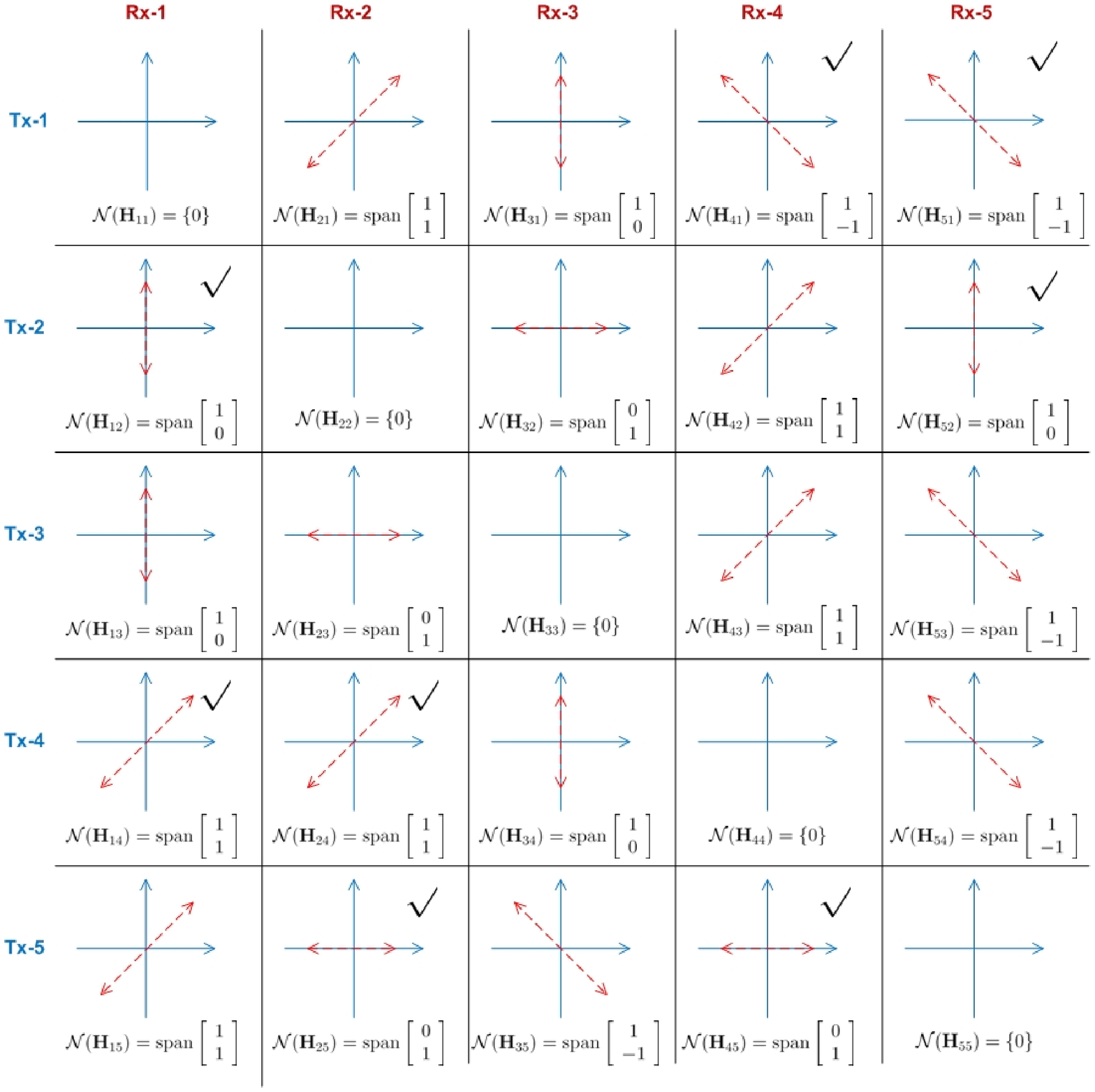}
\caption {Partial connectivity state (Tx side) of a $2\times 2$,
5-pair interference channel. } \label{fig_example}
\end{figure}

Now suppose the network is partially connected such that the channel
matrices of all cross links are rank 1 with null spaces given by the
red arrows in Fig.~\ref{fig_example}. Then we have IA constraints
\eqref{eqn:rank_0}, \eqref{eqn:cs_cross_0} are satisfied under the
following policy: Assign data streams to Tx-Rx pairs 1,2,4,5
($\{1,2,...,K\}$=\{1,2,4,5\}), with precoders
$\mathbf{v}_1=\frac{\sqrt{2}}{2}\left[\begin{array}{c}1\\-1\end{array}\right]$,
$\mathbf{v}_2=\left[\begin{array}{c}1\\0\end{array}\right]$,
$\mathbf{v}_4=\frac{\sqrt{2}}{2}\left[\begin{array}{c}1\\1\end{array}\right]$,
$\mathbf{v}_5=\left[\begin{array}{c}0\\1\end{array}\right]$, and
decorrelators
$\mathbf{u}_1=(\left[\begin{array}{r@{\;\;}r}0&-1\\1&0\end{array}\right]\mathbf{H}_{15}\mathbf{v}_5)^T$,
$\mathbf{u}_2=(\left[\begin{array}{r@{\;\;}r}0&-1\\1&0\end{array}\right]\mathbf{H}_{21}\mathbf{v}_1)^T$,
$\mathbf{u}_4=(\left[\begin{array}{r@{\;\;}r}0&-1\\1&0\end{array}\right]\mathbf{H}_{42}\mathbf{v}_2)^T$,
$\mathbf{u}_5=(\left[\begin{array}{r@{\;\;}r}0&-1\\1&0\end{array}\right]\mathbf{H}_{54}\mathbf{v}_4)^T$.
Since 4 Tx-Rx pairs can have data streams simultaneously, 1 extra
DoF is achieved compared to the fully connected case.

The example above illustrates how partial connectivity can
contribute to network performance gains.  However, as we shall
explain below, there are various technical challenges to exploit the
benefit of partial connectivity for general scenarios.

\subsubsection{\bf The difficulty in exploiting the  benefit of partial connection}
\label{sec:dif}
\begin{itemize}
\item {\bf Interference overlapping versus interference nulling:} Classical interference
alignment schemes \cite{JafarK1} reduce interference dimension by
``overlapping" the interferences from different Txs. However, when
partial connectivity is considered, ``overlapping" interferences is
no longer the only method to reduce interference dimension. Part of
the interference can also be eliminated  by utilizing the null
spaces of the channel states. For instance, in the previous example,
by setting
$\mathbf{v}_1=\frac{\sqrt{2}}{2}\left[\begin{array}{c}1\\-1\end{array}\right]$,
we eliminate the interference from Tx 1 to Rx 4 and 5 by utilizing
the null space of $\mathbf{H}_{41}$ and $\mathbf{H}_{51}$. In
practice, we should dynamically combine these two approaches in
order to better exploit the nulling opportunities as well as
alignment opportunities in a partially connected interference
network. However, a combined design may depend heavily on the
specific realization of the connection topology and a combination
criteria that can work for general connection topologies is not yet
clear.
\item {\bf Freedoms versus constraints:} Another perspective of the technical challenges is
on  the feasibility conditions in quasi-static MIMO interference
networks. In \cite{JafarDf}, a symmetrical MIMO interference system
is \emph{feasible} if and only if the number of freedoms in
transceiver design is no less than the number of independent
constraints induced by interference alignment requirements. In order
to reduce the number of the independent constraints in a partially
connected network, we need to restrict the precoders and
decorrelators to a lower rank subspace. On the other hand,
restricting precoders and decorrelators to a lower rank subspace
reduces the number of the freedoms in the precoders and
decorrelators. The conflict between freedoms in transceiver design
and independent constraints makes the subspace selection very
challenging.
\item{\bf Exponential complexity in checking IA
feasibility conditions:} As revealed in \cite{JafarDf}, checking the
IA feasibility condition requires comparison of freedoms versus
constraints for \emph{every} possible combination of the
interference nulling constraints. This process involves
$2^{K(K-1)}-1$ comparisons, where $K$ is the number of Tx-Rx  pairs.
Such a complexity is intolerable in practice. Hence, a low
complexity algorithm for checking the feasibility condition on a
real-time basis is needed.
 \end{itemize}

\subsection{Dynamic Interference Mitigation Scheme for
a 5-Pair 2$\times$2 Partially
 Connected Interference Network
} \label{sec:example}The following observation is the key insight of
the proposed algorithm:

\emph{Observation: In a partially connected MIMO interference
network, by properly restricting precoders $\mathbf{V}_{m}$ and
decorrelators $\mathbf{U}_{n}$ to a lower rank subspace, we can
eliminate ``many" independent constraints at a cost of only a few
``free variables"  and hence extend the IA feasibility region.}

\begin{figure} \centering
\includegraphics[scale=0.55]{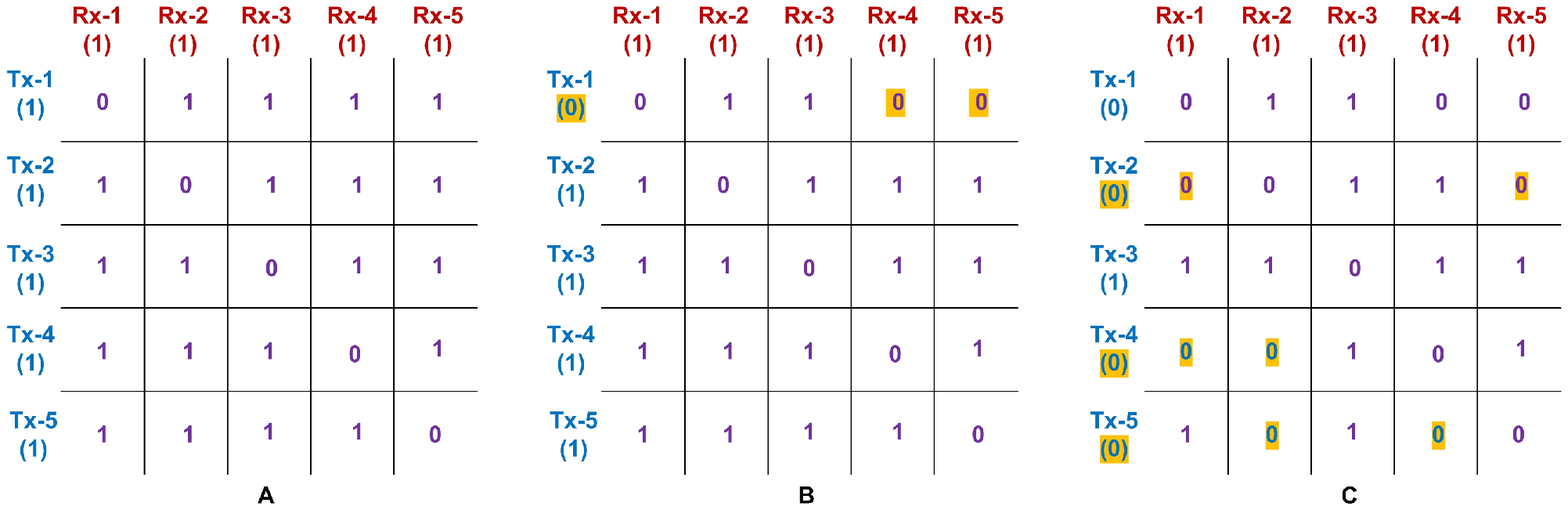}
\caption {The number of the freedoms in precoder and decorrelator
design versus the number of the remaining constraints in the
$2\times 2$, 5-pair interference network before and after subspace
design (Step 1$\sim$3 of the Algorithm in
Section~\ref{sec:example}). The numbers in red and blue denote the
remaining freedoms in the corresponding decorrelators and precoders,
respectively, and the numbers in purple denote the number of the
remaining constraints to null off the interference.}
\label{fig_example_process}
\end{figure}

\begin{figure} \centering
\includegraphics[scale=0.44]{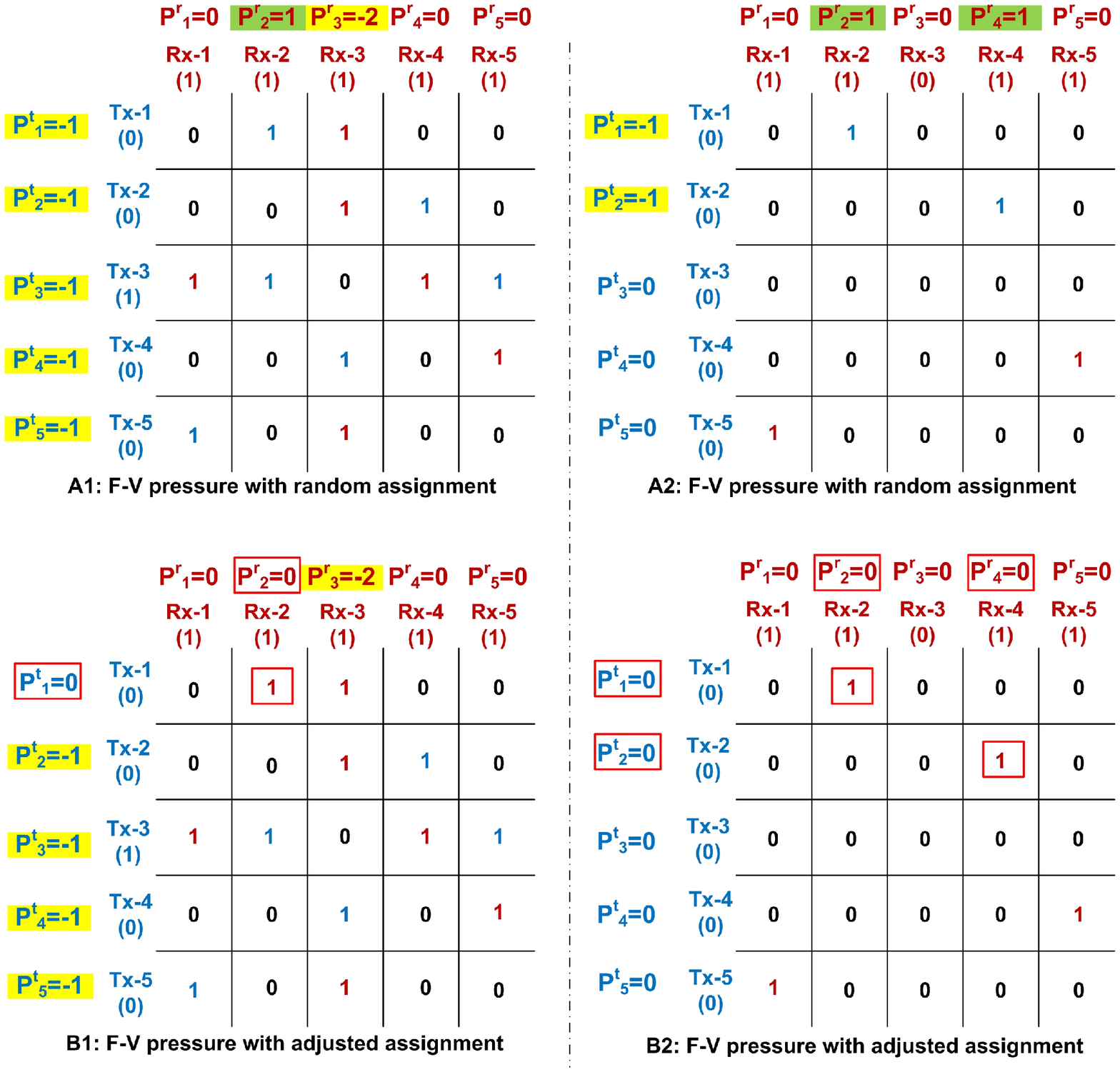}
\caption {The freedom - variable pressure at each node
(i.e. $P^t_n$, $P^r_m$, $n,m\in\{1,2,3,4,5\}$, please refer to
\eqref{eqn:pressure} for mathematical definition) before and after
adjusting constraint assignments (Step 4 of the Algorithm in
Section~\ref{sec:example}). The overloaded nodes and the nodes with
extra freedoms are marked out with yellow and green color,
respectively. The reassignment processes are highlighted using red
boxes.} \label{fig_example_result}
\end{figure}

In this section, we shall use the example of a 5-pair 2$\times$2
partially connected interference network described in
Section\ref{sec:alg_benifit} to illustrate the main ideas of the
proposed scheme.
\begin{itemize}
\item{\bf Step 1 Initialization:} Note that all the 5 direct links
have sufficient rank (rank$(\mathbf{H}_{nn})=2>1$,
$n\in\{1,2...,5\}$), initialize
$\mathbb{D}=\{1,1,1,1,1\}$ to see if the network is
feasible with all Tx-Rx pairs active.

After the initialization, the number of the freedoms in transceiver
design and the number of the IA constraints is illustrated in
Fig.~\ref{fig_example_process}A. The numbers in red and blue denote
the freedoms in the corresponding decorrelators and precoders,
respectively, and the numbers in purple denote the number of the IA
constraints.
\item{\bf Step 2 Find out the common subspaces in partial connectivity:}
As indicated by the``$\surd$" signs in Fig.~\ref{fig_example}, from
Tx 1 to Rx 4 and 5, there is a one dimensional common subspace in
partial connectivity state:
$\mathcal{N}(\mathbf{H}_{41})=\mathcal{N}(\mathbf{H}_{51})=\mbox{span}
\left[\begin{array}{c}1\\-1\end{array}\right]$. Similarly, from Tx 2
to Rx 1 and 5, from Tx 4 to Rx 1 and 2, from Tx 5 to Rx 2 and 4,
there are common null spaces.
\item{\bf Step 3 Select subspaces to reduce the number of IA constraints:}
Set
$\mathbf{v}_1=\left[\begin{array}{c}\frac{\sqrt{2}}{2}\\-\frac{\sqrt{2}}{2}\end{array}\right]$,
we have $\mathbf{H}_{m1}\mathbf{v}_1=\mathbf{0}$, $m\in\{4,5\}$.
Hence, as indicated by the highlight parts in
Fig.~\ref{fig_example_process}B, we reduce 2 constraints at a cost
of 1 freedom. Similarly, as indicated by the highlight parts in
Fig.~\ref{fig_example_process}C, since Tx 2,4,5 each has two cross
links with overlapping null spaces, we can reduce 2 constraints at a
cost of 1 freedom by setting the precoder vectors to be the basis
vectors of the corresponding null spaces.
\item{\bf Step 4 Check the feasibility conditions:} The number of the
remaining freedoms and constraints after step 3 is illustrated in
Fig.~\ref{fig_example_process}C. Randomly assign the
constraints to the corresponding Txs or Rxs (as indicated by the
color of the numbers, deep-blue and deep-red indicate the number is
assigned to the Tx or the Rx, respectively), we get
Fig.~\ref{fig_example_result}A1. In this figure, some nodes are
``overloaded" in the sense that the number of freedoms at this node
minus the number of constraints assigned to this node is negative
(highlighted using yellow) while some nodes still have extra
freedoms (highlighted using green). Hence, as illustrated in
Fig.~\ref{fig_example_result}A2, we can reassign the constraints so
that we can have less overloaded nodes (the changes are highlighted
using red boxes). However, in Fig.~\ref{fig_example_result}A2, there
are still some overloaded nodes while no nodes have extra freedoms,
hence the network is not feasible.
\item{\bf Step 5 Turn off the most ``constraint demanding" stream}:
As illustrated in Fig.~\ref{fig_example_result}A2, if we remove
Tx-Rx  pair 1 from the active set $\{1,2,...,K\}$, we reduce 1
freedom (the freedom in $\mathbf{v}_1$) and 4 constraints (link Tx 1
to Rx 2,3 and Tx 3,5 to Rx 1). Hence, the freedom-constraint gain by
removing Tx-Rx pair 1 is $(-1)-(-4)=3$. Similarly, the
freedom-constraint gains by removing Tx-Rx  pair 2,4,5 are also $3$,
whereas the gain by removing Tx-Rx  pair 3 is $(-2)-(-8)=6$. Since
$6>3$, we remove the stream of Tx-Rx pair 3, i.e. let
$\mathbb{D}=\{1,1,0,1,1\}$ and return to Step 3.
\item{\bf Repeat Step 3$\sim$4:} After removing the stream of Tx-Rx
pair 3, repeat Step 3$\sim$4 and the number of the remaining
freedoms and constraints after subspace selection is illustrated in
Fig.~\ref{fig_example_result}B1. As revealed by
Fig.~\ref{fig_example_result}B2, the IA constraints
can be assigned properly so that no node is overloaded, the network
is feasible. Continue to Step 6.
\item{\bf Step 6 Precoder and decorrelator determination}: Based on
the results given in Steps 1$\sim$5, use the minimum interference
leakage iteration \cite{JafarD1} to determine the precoders and
decorrelators and we get the policy proposed in
Section~\ref{sec:alg_benifit}.\end{itemize}
\begin{Remark}[Low Complexity IA Feasibility Checking in Step 4]  To avoid the
exponential complexity in IA feasibility checking,
 we have proposed a low complexity method,
namely the {\em freedom-constraint assignment} with
worst case complexity $\mathcal{O}(K^3)$ only. Furthermore, we shall
formally prove in Appendix~\ref{pf_thm:proper} that this method is
indeed a necessary and sufficient condition of the IA
feasibility conditions \eqref{eqn:proper_1} in the general
case.~\hfill \IEEEQED
\end{Remark}
\subsection{Dynamic Interference Mitigation Scheme - General Case}
Inspired by the example above, we shall propose a two-stage
algorithm, namely \emph{stream assignment and subspace
determination stage} and {\em precoder / decorrelator determination
stage}. The first stage algorithm (which corresponds to Step
1$\sim$5 in the example illustrated above) determines
the stream assignment pattern
$\mathbb{D}=\{d_1,d_2,...,d_K\}$ and the subspaces for the
precoders $\mathbb{S}^t_n$ and decorrelators $\mathbb{S}^r_m$ for
$n,m\in\{1,2,...,K\}$ based on the partial connectivity state
$\{{\mathcal{N}(\mathbf{H}_{mn})},
{\mathcal{N}^H(\mathbf{H}_{mn})}\}$, $m,n\in\{1,2,...,K\}$.  Based
on the outputs of the stage I algorithm and the channel state, the
stage II algorithm (which corresponds to Step 6 in the example
illustrated above) determines the precoders/decorrelators. In the
next two subsections, we shall elaborate the details of the stage I
and stage II algorithms, respectively.
\subsubsection{Stage I: Stream Assignment and Subspace
Determination} Suppose the row vectors in precoder $\mathbf{V}_{n}$
and the column vectors in decorrelator $\mathbf{U}_{m}$ are
constrained to the linear spaces $\mathbb{S}^t_n
\subseteq\mathbb{C}^{N_t\times 1}$ and $\mathbb{S}^r_m
\subseteq\mathbb{C}^{1\times N_r}$, respectively. Denote
$S^t_n=|\mathbb{S}^t_n|$, $S^r_m=|\mathbb{S}^r_m|$, where
$|\mathbb{X}|$ denotes the rank of linear space $\mathbb{X}$. Then
\eqref{eqn:rank_0} and \eqref{eqn:cs_cross_0} can be rewritten as:
\begin{eqnarray}
\mbox{rank}(\mathbf{U}_{n}\mathbf{H}_{nn}
\mathbf{V}_{n})=\mbox{rank}(\mathbf{U}'_{n}\mathbf{S}^r_n\mathbf{H}_{nn}\mathbf{S}^t_n
\mathbf{V}'_{n})&=&d_n\label{eqn:cs_rank_1}
\\\mathbf{U}_{m}\mathbf{H}_{mn}\mathbf{V}_{n}=\mathbf{U}'_{m}\mathbf{S}^r_m\mathbf{H}_{mn}
\mathbf{S}^t_n\mathbf{V}'_{n}&=&\mathbf{0}\label{eqn:cs_cross_1}
\end{eqnarray}
where the column vectors of the $N_t\times S^t_n$ matrix
$\mathbf{S}^t_n$ and the $S^r_m\times N_r$ matrix $\mathbf{S}^r_m$
span the spaces $\mathbb{S}^t_n$ and $\mathbb{S}^r_m$, respectively.
Note that the \emph{constrained} precoder $\mathbf{V}'_{n}$ and the
\emph{constrained} decorrelator $\mathbf{U}'_{m}$ are $S^t_n\times
d_n$ and $d_m \times S^r_m$ matrices, respectively.

Before we elaborate the details of the stage I processing, we shall
first define the notion of a {\em proper MIMO interference system}.
\begin{Def}[Proper
MIMO Interference Systems for the General Case]\label{def:proper}
The MIMO interference system is proper for interference alignment
if:
\begin{eqnarray}
\nonumber
&&\sum_{(n,m)\in\mathbb{G}}\min\left(d_m,|\mathbb{S}^r_m\cap(\mathcal{N}^H(\mathbf{H}_{mn}))^\bot|\right)\cdot\min\left(
|\mathbb{S}^t_n\cap\mathcal{N}(\mathbf{H}_{mn})^\bot|, d_n \right)
\\&& \le
\sum_{n\in\mathbb{G}_T}d_n(S^t_n-d_n)+\sum_{m\in\mathbb{G}_R}d_m(S^r_m-d_m)\label{eqn:proper_1}
\end{eqnarray}
$\forall \mathbb{G}_T,\mathbb{G}_R\subseteq \{1,2,...,K\}$,
$\mathbb{G}= (\mathbb{G}_R\times\mathbb{G}_T)\backslash \{(n,n)$
$,n\in\mathbb{G}_R \cap \mathbb{G}_T\}$, where ``$\times$" denotes
Cartesian product.~\hfill \IEEEQED
\end{Def}
\begin{Remark}[Physical Meaning of Definition~\ref{def:proper}]
As proved in Appendix~\ref{sec:proper}, the left hand side of
\eqref{eqn:proper_1} represents the total number of interference
alignment constraints of the links from a Tx in set $\mathbb{G}_n$
to a Rx in set $\mathbb{G}_m$, and the first and second term on the
right hand side of \eqref{eqn:proper_1} represent the sum of the
free variables in $\mathbf{V}'_n$, $n\in\mathbb{G}_n$ and
$\mathbf{U}'_m$, $m\in\mathbb{G}_m$, respectively. Hence,
\eqref{eqn:proper_1} means that for any subset of Tx-Rx combination
$\mathbb{G}_n\oplus \mathbb{G}_m\subseteq
\{1,2,...,K\}\oplus\{1,2,...,K\}$, the number of constraints is no
more than the number of free variables.
 ~\hfill \IEEEQED
\end{Remark}

The main steps of the Stage I processing algorithm for the general
$K$-pair partially connected interference network is
illustrated below. Steps 1$\sim$5 below corresponds to the Steps
1$\sim$5 for the 5-Pair example.
\begin{Alg}[Stream Assignment and Subspace Determination]\label{alg:offline}
\begin{itemize}
\item[]
\item{\bf Step1 Initialization:} Initialize the
number of stream assigned to each Tx-Rx pair to be the minimum of
the rank of the direct link and the number of streams  claimed by
this Tx-Rx pair, i.e. $d_n=\min(\mbox{rank
}(\mathbf{H}_{nn}),d^{\max}_n)$, $\forall n\in\{1,2,...,K\}$.
\item{\bf Step2 Calculate the common null spaces\footnote{The worst case
complexity of this step is $\mathcal{O}(K2^{K'-1})$, where
$K'=\max_n(\max(|\mathbb{K}^t_n|,|\mathbb{K}^r_n|))$. In practice,
due to path loss, $K'$ usually does not scale with $K$. Hence, the
$2^{K'-1}$ term is only a moderate constant which does not scale
with the size of the network in most of the interesting scenarios.} :}
For every Tx-n, $n\in\{1,2,...,K\}$, calculate the
common subspaces of the null spaces of the cross links from this Tx
within the effective subspace of the direct link , i.e.
$\mathcal{N}(\{\mathbf{H}_{mn}:m\in\mathbb{K}_{sub}\})=\left(\cap_{m\in\mathbb{K}_{sub}}
\mathcal{N}(\mathbf{H}_{mn})\right)\cap(\mathcal{N}(\mathbf{H}_{nn}))^\perp$,
$\mathbb{K}_{sub}\subseteq\{1,2,...,K\}$ as follows:
\begin{itemize}
\item Denote $\mathbb{K}^t_{n}=\{m:m\in\{1,2,...,K\},
\mathcal{N}(\mathbf{H}_{mn})\neq\mathbb{C}^{N_t\times1}(\mbox{i.e. }
\mathbf{H}_{nm}\neq \mathbf{0})\}$, Initialize
$\mathcal{N}(\emptyset)=(\mathcal{N}(\mathbf{H}_{nn}))^\perp$,
$\mathcal{N}(\{\mathbf{H}_{mn}\})=\mathcal{N}(\mathbf{H}_{mn})\cap
(\mathcal{N}(\mathbf{H}_{nn}))^\perp$, $\forall m\in
\mathbb{K}^t_{n}$, and subset cardinality parameter $C=2$.
\item For every $\mathbb{K}_{sub}\subseteq\mathbb{K}^t_{n}$ with
$|\mathbb{K}_{sub}|=C$, if all the subsets of $\mathbb{K}_{sub}$
with cardinality $(C-1)$ are not $\{0\}$, calculate
$\mathcal{N}(\{\mathbf{H}_{mn}:m\in\mathbb{K}_{sub}\})=\mathcal{N}(\{\mathbf{H}_{mn}:
m\in\mathbb{K}_{sub}\backslash\{m'\}\}\cap
\mathcal{N}(\mathbf{H}_{m'n})$, where $m'$ is an arbitrary element
in $\mathbb{K}_{sub}$. Update $C=C+1$. Repeat this process until
$\mathcal{N}(\{\mathbf{H}_{mn}:m\in\mathbb{K}_{sub}\})=\{0\}$,
$\forall\mathbb{K}_{sub}\subseteq\mathbb{K}^t_{n}$ with
$|\mathbb{K}_{sub}|=C$ or $C=|\mathbb{K}^t_{n}|$.
\item For every $\mathbb{K}_{sub}\subseteq\mathbb{K}^t_{n}$ with
$\mathcal{N}(\{\mathbf{H}_{mn}:m\in\mathbb{K}_{sub}\})\neq \{0\}$,
set
$\mathcal{N}(\{\mathbf{H}_{mn}:m\in\mathbb{K}_{sub}\cup(\{1,2,...,K\}\backslash\mathbb{K}^t_{n})\})
=\mathcal{N}(\{\mathbf{H}_{mn}:m\in\mathbb{K}_{sub}\})$.
\end{itemize}
For every Rx-m, $m\in\{1,2,...,K\}$, calculate
$\mathcal{N}^H(\{\mathbf{H}_{mn}:n\in\mathbb{K}_{sub}\})=(\cap_{n\in\mathbb{K}_{sub}}
\mathcal{N}^H(\mathbf{H}_{mn}))\cap(\mathcal{N}^H(\mathbf{H}_{mm}))^\perp$,
$\mathbb{K}_{sub}\subseteq\{1,2,...,K\}$ using a similar process.
\item{\bf Step3 Design Subspace constraints $\mathbb{S}^t_{n}$ and $\mathbb{S}^r_{m}$:}
For every Tx $n$,  $n\in\{1,2,...,K\}$, generate a
series of potential subspace constraints $\mathbb{S}^t_{n}(d),
d\in\{d_n,d_n+1,...,N_t\}$ with $|\mathbb{S}^t_{n}(d)|=d$,
 based on the principe that a subspace which has
higher null space ``weight"\footnote{The weight of
$\mathcal{N}(\{\mathbf{H}_{mn}:m\in\mathbb{K}_{sub}\})$ is
$\sum_{m\in\mathbb{K}_{sub}}d_m$. From the left hand side of
\eqref{eqn:proper_1}, this weight is the maximum number of IA
constraints that one can mitigate by selecting a one dimensional
subspace in
$\mathcal{N}(\{\mathbf{H}_{mn}:m\in\mathbb{K}_{sub}\})$.} is
selected with higher priority. Choose the subspace constraint
$\mathbb{S}^t_{n}$ from the potential subspace constraints:
$\mathbb{S}^t_{n}=\mathbb{S}^t_{n}(d^*)$, where:
 \begin{eqnarray}
\nonumber
d^*&=&\arg\max_{d\in\{d_n,d_n+1,...,N_t-|\mathcal{N}(\mathbf{H}_{nn})|\}}
 d_n(d-d_n)-
 \\&&\sum_{m\neq
n}^{\{1,2,...,K\}}\min(d_m,|({\mathcal{N}^H(\mathbf{H}_{mn})})^\bot|)\min
(|\mathbb{S}^t_{n}(d)\cap\mathcal{N}(\mathbf{H}_{mn})^\bot|,d_n).
\label{eqn:d_t}\end{eqnarray}

For every Rx {$m$}, $m\in\{1,2,...,K\}$, use a similar
process to generate $\mathbb{S}^r_{m}(d)$,
$d\in\{d_m,d_m+1,...,N_r\}$ and set
$\mathbb{S}^r_{m}=\mathbb{S}^r_{m}(d^*)$, where
\begin{eqnarray}\nonumber d^*&=&\arg\max_{d\in\{d_m,d_m+1,...,N_r-|\mathcal{N}^H(\mathbf{H}_{mm})|\}}
d_m(d-d_m)-
\\&&\sum_{n\neq
m}^{\{1,2,...,K\}}\mbox{min}(d_m,|\mathbb{S}^r_{m}(d)\cap(\mathcal{N}^H(\mathbf{H}_{mn}))^\bot|)
\min
(|\mathbb{S}^t_{n}\cap({\mathcal{N}(\mathbf{H}_{mn})})^\bot|,d_n
).\label{eqn:d_r}
\end{eqnarray}
\item{\bf Step4 Low complexity feasibility checking:}
Denote $v^t_n,v^r_m$, $n,m\in\{1,2,...,K\}$ as the number of the
freedoms at Tx {$n$} and Rx {$m$},
respectively. Set $v^t_n=d_n(|\mathbb{S}^t_{n}|-d_n)$,
$v^r_m=d_m(|\mathbb{S}^r_{m}|-d_m)$. Denote $c_{mn}$, $n\neq
m\in\{1,2,...,K\}$ as the number of constraints required to
eliminate the interference from the Tx {$n$} to the Rx
{$m$}. Set
$c_{mn}=\min\left(d_m,|\mathbb{S}^r_{m}\cap({\mathcal{N}^H(\mathbf{H}_{mn})})^\bot|\right)$
$
\min\left(d_n,|\mathbb{S}^t_{n}\cap{\mathcal{N}(\mathbf{H}_{mn})}^\bot|\right)
$ and $c_{mm}=0$, $\forall n\neq m\in\{1,2,...,K\}$. Use
freedom-constraint assignment to check if the system is proper
(Please refer to Appendix~\ref{sec:alg_detail} for details.)
 If the network is not proper, go to
Step 5. Otherwise, let
$\mathbb{D}^*=\{d^*_1,d^*_2,...,d^*_K\}=\mathbb{D}$,
$\mathbb{S}^{t*}_n=\mathbb{S}^{t}_n$,
$\mathbb{S}^{r*}_m=\mathbb{S}^{r}_m$, $\forall n,m\in\{1,2,...,K\}$,
and exit the algorithm.
\item{\bf Step5 :} {
  Update $\mathbb{D}=\{d_1,d_2,...d_{\tilde{n}}-1,...,d_K\}$ and go back to Step
3, where $\tilde{n}$ is given by
\begin{eqnarray}\tilde{n}=\arg\max_{n\in\{1,2,...,K\}}\left(\sum_{m=1}^{K}(c_{mn}+c_{nm}-c'_{mn}-c'_{nm})-
(v^t_n+v^r_n-{v^t_n}'-{v^r_n}')\right)\label{eqn:selectn}\end{eqnarray}
where $\{{v^t_n}',{v^r_n}'\}$ and $\{c'_{mn},c'_{nm}\}$,
$m\in\{1,2,...,K\}$ are the number of freedoms and IA constraints
under updated subspace constraints ${\mathbb{S}^t_n}'$ and
${\mathbb{S}^r_n}'$ given by \eqref{eqn:d_t} and \eqref{eqn:d_r}
with $d'_n=d_n-1$, respectively.}
\end{itemize}
\end{Alg}
\begin{Remark}[stream assignment and subspace design
criterion in Stage I Algorithm] As revealed in \cite{JafarDf}, the
IA feasibility condition is the major limitation of the DoF
performance achieved by MIMO interference networks. Hence, in order
to enhance the network DoF performance, subspace constraints
$\{\mathbb{S}^{t*}_n,\mathbb{S}^{r*}_m\}$ and stream assignment
$\mathbb{D}^*_n$ are designed to alleviate the IA feasibility
condition as much as possible.
\begin{itemize}
\item Recall \eqref{eqn:numv1}, \eqref{eqn:numc1}, we can see that
\eqref{eqn:d_t}, \eqref{eqn:d_r} choose
 the dimension of subspace constraints $d^*$ to maximize the
 difference between the number of freedoms in precoder (or decorrelator) design
 minus the number of IA constraints endued by the Tx (or Rx).
 {\item  The right hand side of \eqref{eqn:selectn}
represents the number of IA constraint - freedom in transceiver
design saved by removing one stream from Tx-Rx  pair $n$. The higher
this number, the more ``constraint demanding" this stream is. Hence,
it should be removed first so that the network can become IA
feasible easier.}~\hfill \IEEEQED
 \end{itemize}
\end{Remark}
{\begin{Lem}[Property of the low complexity IA
feasibility checking] A partially connected MIMO interference
network with stream assignment $\mathbb{D}$ and potential signal
subspace $\{\mathbb{S}^{t}_n,\mathbb{S}^{r}_m \}$ policy is proper
(i.e. satisfies \eqref{eqn:proper_1}) if and only if it can pass the
low complexity IA feasibility checking in
Appendix~\ref{sec:alg_detail}. Moreover, the worst case complexity
of the proposed checking scheme is
$\mathcal{O}(K^3)$.\label{thm:proper}
\end{Lem}
\proof Please refer to Appendix~\ref{pf_thm:proper} for the proof.
\endproof}
From Lemma~\ref{thm:proper}, and Definition~\ref{def:proper}, we
have the following theorem:
\begin{Thm}[Property of $\{\mathbb{S}^{t*}_n,\mathbb{S}^{r*}_m \}$ and $\mathbb{D}^*$]
\label{thm:proper2} Under Assumption~\ref{asm:channel}, the
potential signal subspace $\{\mathbb{S}^{t*}_n,\mathbb{S}^{r*}_m
\}$, together with the stream assignment pattern  $\mathbb{D}^*$
from Stage~\ref{alg:offline} form a \emph{proper} system.
\end{Thm}

\subsubsection{Stage II: Precoder and Decorrelator Determination}

In this section, we shall elaborate the stage II processing, which
 determines the precoders
$\{\mathbf{V}_n\}$ and the decorrelators $\{\mathbf{U}_m\}$ under
the subspace constraints determined in stage I. Specifically, since
the subspace constraint can be represented by the constrained
precoder and decorrelator as in \eqref{eqn:cs_rank_1},
\eqref{eqn:cs_cross_1}, the precoder and decorrelator design is
given by the following optimization objective which minimizes the
total interference leakage power in the network:

$\min_{\mathbf{U}'_m,\mathbf{V}'_n}\sum_{n=1,d^*_n>0}^{K}\sum_{m=1,\neq
n\atop
d^*_m>0}^{K}\frac{P_n}{d^*_n}\mbox{trace}\left((\mathbf{U}'_{m}\mathbf{S}^{r*}_m\mathbf{H}_{mn}
\mathbf{S}^{t*}_n\mathbf{V}'_{n})^H(\mathbf{U}'_{m}\mathbf{S}^{r*}_m\mathbf{H}_{mn}
\mathbf{S}^{t*}_n\mathbf{V}'_{n})\right)$.

The following algorithm is guaranteed to converge to a local optimum
\cite{JafarD1}.

\begin{Alg}[Precoder and Decorrelator Determination]\label{alg:online} Given $\{\mathbb{D}^*,
\mathbb{S}^{t*}_n, \mathbb{S}^{r*}_m\}$, determined by Stage I:
\begin{itemize}
\item{\bf Step1 Initialization :} Denote
$\mathbf{S}^r_m$ and $\mathbf{S}^t_n$ as the structure matrices for
decorrelators and precoders:
$\mathbf{U}_{m}=\mathbf{U}'_{m}\mathbf{S}^r_m$,
$\mathbf{V}_{n}=\mathbf{S}^t_n \mathbf{V}'_{n}$. Set
$\mathbf{S}^{t*}_n$ and $\mathbf{S}^{r*}_m$ to be the aggregation of
the basis vectors in $\mathbb{S}^t_n$ and $\mathbb{S}^r_m$,
respectively. Randomly generate $\mathbf{V}'_n$.
\item{\bf Step2 Minimize interference leakage at the receiver
side:} At each Rx {$m$}, such that $d^*_m>0$, update
$\mathbf{U}'_m$: $\mathbf{u}'_m(d)=\left(\nu_d\left[\sum_{n=1,\neq
m\atop d^*_n>0}^{
K}\frac{P_n}{d^*_n}(\mathbf{S}^{r*}_m\mathbf{H}_{mn}
\mathbf{S}^{t*}_n\mathbf{V}'_{n})(\mathbf{S}^{r*}_m\mathbf{H}_{mn}
\mathbf{S}^{t*}_n\mathbf{V}'_{n})^H\right]\right)^H$, where
$\mathbf{u}'_m(d)$ is the $d$-th row of $\mathbf{U}'_m$,
$\nu_d[\mathbf{A}]$ is the eigenvector corresponding to the d-th
smallest eigenvalue of $\mathbf{A}$,
$d\in\{1,2,...,{d^*_m}\}$.
\item{\bf Step3 Minimize interference leakage at the transmitter side:}
At each Tx {$n$} such that $d^*_n>0$,  update
$\mathbf{V}'_n$: $\mathbf{v}'_n(d)=\nu_d\left[\sum_{m=1,\neq n\atop
d^*_m>0}^{
K}\frac{P_n}{d^*_n}(\mathbf{U}'_{m}\mathbf{S}^{r*}_m\mathbf{H}_{mn}
\mathbf{S}^{t*}_n)^H(\mathbf{U}'_{m}\mathbf{S}^{r*}_m\mathbf{H}_{mn}
\mathbf{S}^{t*}_n)\right]$, where $\mathbf{v}'_m(d)$ is the $d$-th
column of $\mathbf{V}'_m$, $d\in\{1,2,...,{d^*_n}\}$.
\item [] Repeat Step 2 and 3 until $\mathbf{V}'_n$ and $\mathbf{U}'_m$
converges. Set $\mathbf{V}^*_n=\mathbf{S}^t_n\mathbf{V}'_n$ and
$\mathbf{U}^*_m=\mathbf{U}'_m\mathbf{S}^r_m$, $\forall
n,m\in\{1,2,...,K\}$.
\end{itemize}
\end{Alg}

\begin{Remark}[Backward Compatibility
 of the Proposed Scheme] When the system is fully connected
 (i.e. $\mathcal{N}(\mathbf{H}_{mn})=\mathcal{N}^H(\mathbf{H}_{mn})=\{0\}$, $\forall
 n,m\in\{1,2,...,K\}$), it is easy to check in Algorithm 1,
  $\mathbb{S}^{t*}_n=\mathbb{C}^{N_t\times
 1}$ and $\mathbb{S}^{r*}_m=\mathbb{C}^{1\times
N_r}$; this means that in fully connected quasi-static MIMO interference
 networks, the proposed scheme reduces to the conventional IA
 schemes proposed in \cite{JafarDf} and \cite{JafarD1}. When $N_t>
 N_r$, the algorithm shall first utilize the null spaces on the Tx side and
 design $\mathbb{S}^{t*}_n$ to null
 off part of the interference, which is similar to \cite{JafarX1}.
 However, given a general partial connectivity topology $\{\mathcal{N}(\mathbf{H}_{mn}),\mathcal{N}^H(\mathbf{H}_{mn})\}$,
 the algorithm generalizes the conventional interference alignment
 by dynamically combining the interference alignment
 and interference nulling approaches.
 ~\hfill \IEEEQED
\end{Remark}

\section{Performance Analysis}
{In this section, we shall derive the analytical DoF
performance achieved by the proposed scheme in a symmetrical
partially connected MIMO interference network.} In fact, although
the algorithm itself applies to general typologies, analyzing such
cases can be prohibitively complicated. Since there are too many
parameters in the general partial connectivity parameters
$\{\mathcal{N}(\mathbf{H}_{mn}),\mathcal{N}^H(\mathbf{H}_{mn})\}$,
$m,n\in\{1,2,...,K\}$ in Section~\ref{sec:model}, we shall focus on
a symmetrical {$K$}-pair partially connected MIMO
interference network in which the partial connectivity is induced by
both the path loss and the local scattering.

\begin{Def}[Symmetrical Partially Connected MIMO Interference Channels]
\label{def:model} Consider a {$K$}-pair partially
connected MIMO interference network with the following
configuration. Each Tx has $N_t$ antennas and each Rx has $N_r$
antennas. Each Tx-Rx  claims $d^{\max}_n=d_f$ data streams, $\forall
n\in\{1,2,...,K\}$. The partial connectivity states (elaborated
below) are expressed in terms of three key parameters $L$, $E_1$ and
$E_2$, which characterize the connection density, the rank of the
direct links and the rank of the cross links, respectively. Please
refer to Appendix~\ref{sec:reason} for the details.
\begin{eqnarray}
\mathcal{N}(\mathbf{H}_{mn})=\left\{\begin{array}{l}\left(\mbox{span}(\mathbf{e}_{N_t}(q))\right)^\perp,
q\in\mathbb{E}_1 \mbox{ if: } m=n
\\ \left(\mbox{span}(\mathbf{e}_{N_t}(q))\right)^\perp,
q\in\mathbb{E}_2(n-m) \mbox{ if: } {0<|n-m|\le L \mbox{
or } |n-m|\ge K-L}
\\\mathbb{C}^{N_t \times 1} \mbox{
otherwise}\end{array}\right.\label{eqn:NT_a}
\\\mathcal{N}^H(\mathbf{H}_{mn})=\left\{\begin{array}{l}
\left(\mbox{span}((\mathbf{A}_R\mathbf{h}^{ang}_{mn}(q))^H)\right)^{\perp},
q\in\mathbb{E}_1 \mbox{ if: } m=n
\\ \left(\mbox{span}((\mathbf{A}_R\mathbf{h}^{ang}_{mn}(q))^H)\right)^{\perp},
q\in\mathbb{E}_2(n-m)\mbox{ if: }\begin{array}{l} {0<|n-m|\le L
\mbox{ or }|n-m|}
\\{\ge
K-L}\end{array}
\\\mathbb{C}^{1 \times N_r} \mbox{
otherwise}\end{array}\right.\label{eqn:NE_a}
\end{eqnarray}
where $\mathbf{H}^{ang}_{mn}$, $\mathbf{A}_R$ and
$\mathbf{e}_{N_t}(q)$ are defined in \eqref{eqn:ant_vir},
\eqref{eqn:ar} and \eqref{eqn:e}, respectively.
$\mathbf{h}^{ang}_{mn}(s)\in\mathbb{C}^{N_r\times 1}$ is the $s$-th
column of $\mathbf{H}^{ang}_{mn}$. $\mathbb{E}_1$ and
$\mathbb{E}_2(n-m)$ are subsets of $\{1,2,...,N_t\}$, with
$|\mathbb{E}_1|=E_1$, $|\mathbb{E}_2(n-m)|=E_2$, $\forall
n,m$.~\hfill \IEEEQED
\end{Def}
\begin{Thm}[Performance of the Partially Connected {$K$}-pair MIMO Systems]
The proposed algorithm could achieve
$\mathbb{D}^*=\{d_f,d_f,...,d_f\}$ if \label{thm:per1}
\begin{eqnarray}
d_f\le\max\left(\frac{E_1+\min(E_1,N_r)}{\min(K-1,2L)+2},
\frac{\min(E_1,N_r)}{\min(K-1,2L)\frac{E_2}{N_t}+1}\right).
\label{eqn:result1}
\end{eqnarray}
\end{Thm}
\proof Please refer to Appendix~\ref{pf_thm:per1} for the proof.
\endproof

\begin{Remark}[Interpretation of the Results] \label{remark:result} Note that the total DoF of the system
is given by $Kd_f$, using \eqref{eqn:result1}, the system can
achieve a total DoF up to:
\begin{eqnarray}K{\left\lfloor\max\left(\frac{E_1+\min(E_1,N_r)}{\min(K-1,2L)+2},
\frac{\min(E_1,N_r)}{\min(K-1,2L)\frac{E_2}{N_t}+1}\right)\right\rfloor}
\label{eqn:result2}\end{eqnarray}
 where the first term and second term in the ``max" operation
 are contributed by restricting the precoders and decorrelators
 in the subspaces $\mathbb{S}^{t*}_n$ and $\mathbb{S}^{r*}_n$ obtained in the stage I algorithm.
 In the following, we shall elaborate various insights regarding
 how the partial connectivity affects the gain of the system.

\begin{itemize}
\item {\bf The gain due to partial connection:} In a {$K$}-pair fully connected
quasi-static MIMO interference channel, the system sum DoF is upper
bounded by $\frac{K(N_t+N_r)}{K+1}$. The partial connectivity
improves this bound in two aspects: 1) \emph{Gain due to path loss}:
As path loss limits the maximum number of Rxs that each Tx may
interfere, the total DoF of the system can grow on $\mathcal{O}(K)$;
2) \emph{Gain due to spatial correlation}: When the spatial
correlation in the cross link is strong (i.e. small $E_2$), a
$\frac{N_t}{E_2}$ factor gain can be further observed.
\item {\bf Connection density versus system performance:} For large $K$, the
DoF of the system \eqref{eqn:result2} scales with
$\sim\mathcal{O}(\frac{1}{L})$, which shows that the network density is
always a first order constraint on the system DoF.
\item {\bf Rank of the cross links versus system performance:}
The system sum DoF \eqref{eqn:result2} is a
{(non-strictly)} decreasing function of $E_2$, which
means system sum DoF grows when the rank of the cross links
decrease. When $E_2=0$, the achievable DoF in \eqref{eqn:result2} is
reduced to: $K\min(E_1,N_r)=K\mbox{rank}(\mathbf{H}_{nn})$, which
means that all Tx-Rx  pairs are using all the dimensions of the
direct link for transmission.
\item {\bf Rank of the direct links versus system performance:}
The system sum DoF \eqref{eqn:result2} is a
{(non-strictly)} increasing function of $E_1$, which
means that the system sum DoF increases when the rank of the direct
links increase. High rank direct links help to increase system
performance from two aspects: 1) They increase the DoF upper bound
that Tx-Rx pairs may achieve. 2) They increase the maximum number of
free variables in the precoders and the decorrelators.
\item {\bf Backward compatibility with previous results:} {When the network is fully connected}, i.e. $E_1=E_2=N_t$ and $L\ge[\frac{K}{2}]$,
the inequality in \eqref{eqn:result1} is reduced to:
$d_f\le\frac{N_t+N_r}{K+1}$, which is consistent with the results in
\cite{JafarDf}.~\hfill \IEEEQED
\end{itemize}
\end{Remark}

\section{Simulation Results}
\begin{figure} \centering
\includegraphics[scale=0.35]{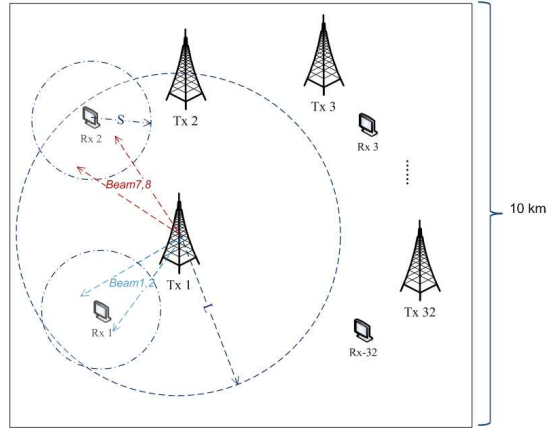}
\caption {An illustration of a randomized partially connected MIMO
interference network. In this illustration, $K=32$, the Txs and Rxs
uniformly distribute in a $10km\times 10km$ square. $L$ is the
maximum distance that a Tx can interfere and $S$ is the radius of
the local scattering.} \label{fig_random}
\end{figure}

In this section, we shall illustrate the performance of the proposed
scheme by simulation. To better illustrate how physical parameters
such as the path loss and the scattering environment affect system
performance, we consider the following simulation setup based on a
randomized MIMO interference channel.
\begin{Def}[Randomized Partially Connected MIMO Interference Channels]
\label{def:model2} We have 32 Tx-Rx  pairs distributed uniformly in
a $10km\times10km$ square as illustrated in Fig.~\ref{fig_random}.
Each node has 12 antennas.\footnote{We choose relatively large
number of Tx-Rx pairs and antennas so that we can have a smooth
system performance variation w.r.t. to partial connectivity
parameters of the network such as $L$ and $S$.} Each Tx-Rx  pair is
trying to deliver 2 data streams. Each Tx is transmitting with power
$P$. Denote $D_{mn}$ as the distance between the Tx
{$n$} and Rx {$m$}. The partial
connectivity is contributed the following factors:
\begin{itemize}
\item {\bf Path loss effect:}
If $D_{mn} > L$, we assume the channel from the Tx
{$n$} to the Rx {$m$} is not connected
($\mathbf{H}_{mn} = \mathbf{0}$).
\item {\bf Local scattering effect:}
If $D_{mn} \le L$, then due to \emph{local scattering}, the angular
domain channel states $\mathbf{H}^{ang}_{mn}=\{h_{mn}(p,q)\}, p,q\in
\{1,2,...,12\}$, has the following property: $h_{mn}(p,q)=0$ if $q$
satisfies \eqref{eqn:virtual_1}, where $S$ is the radius of the
local scattering otherwise $h_{mn}(p,q)\sim \mathcal{CN}(0,1)$.
\end{itemize}
As a result, the partial connectivity parameters for the randomized
model is given by
\begin{eqnarray}
\mathcal{N}(\mathbf{H}_{mn})&=&\left\{\begin{array}{l}\mbox{span}(\mathbf{e}_{N_t}(q)),
q\in\mathbb{Q}_{mn} \mbox{ if: } D_{mn}\le L
\\\mathbb{C}^{12 \times 1} \mbox{ otherwise.}\end{array}\right.
\\ \mathcal{N}^H(\mathbf{H}_{mn})&=&\left\{\begin{array}{l}\left(\mbox{span}((\mathbf{A}_R\mathbf{h}^{ang}_{mn}(q))^H)\right)^{\perp},
q\not\in\mathbb{Q}_{mn} \mbox{ if: } D_{mn} \le L
\\\mathbb{C}^{1 \times 12} \mbox{ otherwise.}\end{array}\right.
\end{eqnarray}
where $\mathbf{H}^{ang}_{mn}$, $\mathbf{A}_R$ and
$\mathbf{e}_{N_t}(q)$ are defined in \eqref{eqn:ant_vir},
\eqref{eqn:ar} and \eqref{eqn:e}, respectively.
$\mathbf{h}^{ang}_{mn}(s)\in\mathbb{C}^{N_r\times 1}$ is the $s$-th
column of $\mathbf{H}^{ang}_{mn}$, and $\mathbb{Q}_{mn}$ is the set
of all the column indices $q\in\{1,2,...,12\}$ that satisfies
\eqref{eqn:virtual_1}. Note that $\mathbb{Q}_{mn}$ is a random set
with randomness induced by the random positions of the Tx
{$n$} and Rx {$m$}.~\hfill \IEEEQED
\end{Def}
\begin{Remark}[Physical Meaning of the Parameters in Definition~\ref{def:model2}]
There are two parameters in Definition~\ref{def:model2}, $L$ and
$S$. As as illustrated in Fig.~\ref{fig_random}, $L$ is the maximum
distance that a Tx can interfere (e.g. the big circle centered at Tx
1 in the figure) and hence reflects the \emph{connection density} of
the network.  $S$ is the radius of the local scattering, from
\eqref{eqn:virtual_1}, if the direction of a beam from the Tx does
not overlap with the local scattering area of a Rx, it cannot be
received by the Rx. (e.g. The local scattering area for Rx 1 is the
small circles centered at Rx 1. Beam 7, 8 cannot be received by Rx 1
as their direction does not overlap with this circle.) Hence, $S$
controls the rank (spatial correlation level) of the non-zero
channels matrices. Larger $S$ corresponds to higher rank
  channel matrices.~\hfill \IEEEQED
\end{Remark}

The proposed interference mitigation scheme is compared with 5
reference baselines below:
\begin{itemize}
\item{\bf{ Conventional interference alignment} (Baseline 1):}
The system directly adapts the precoder-decorrelator iteration
proposed in \cite{JafarD1}.
\item{\bf{ Maximum rank signal subspace}  (Baseline 2):}
{Each node selects a maximum rank  subspace constraint,
i.e. set $d^*$ to be $N_t-|\mathcal{N}(\mathbf{H}_{nn})|$ and
$N_r-|\mathcal{N}^H(\mathbf{H}_{mm})|$ in \eqref{eqn:d_t} and
\eqref{eqn:d_r}, respectively in Stage I,} then uses Stage II to
determine the precoders and decorrelators.
\item{\bf{ Minimum rank signal subspace}  (Baseline 3):}
{Each node selects a minimum rank  subspace constraint,
i.e. set $d^*$ to be $d_n$ and $d_m$ in \eqref{eqn:d_t} and
\eqref{eqn:d_r}, respectively in Stage I,} in the stage I algorithm,
then use the stage II algorithm to determine the precoders and
decorrelators.
\item {\bf TDMA (Baseline 4)}  refers to the case where the Tx-Rx  pairs use
time division multiple access to avoid all interference.
\item{\bf Isotropic transmission (Baseline 5)}  refers to the case where the
Tx and Rx  sends and receives the data streams with random precoders
and decorrelators without regard of the channel information.
\end{itemize}

\subsection{Performance w.r.t. SNR}

\begin{figure} \centering
\includegraphics[scale=0.5]{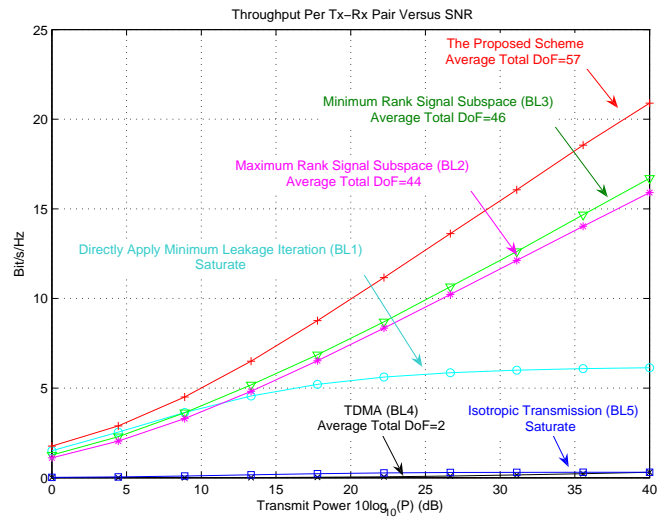}
\caption {Throughput per Tx-Rx  pair versus SNR for the proposed
algorithm (and the 5  baselines) in a randomized partially connected
MIMO interference channel. The parameters are given by $L=5km$ and
$S=3km$.} \label{fig_perf_power}
\end{figure}

Fig.~\ref{fig_perf_power} illustrates the throughput per Tx-Rx pair
versus SNR ($10\log_{10}(P)$). Here $L=5km$ and $S=3km$.
Conventional interference alignment  (BL 1) saturates in the high
SNR region as traditional IA is infeasible in this dense network.
Both the proposed scheme and the Maximum/Minimum signal subspace
methods (BL 2 and 3) can achieve throughputs that grow linearly with
SNR since the on/off selection in the stage I algorithm guarantees
that the system is feasible for IA. However, the proposed scheme
achieves much higher DoF ({$\approx57$}) than BL 2
({$\approx44$}) and 3 ({$\approx46$}),
illustrating the importance of carefully designing the signal
subspaces. Comparison of the proposed scheme and BL 1 shows that
introducing subspace constraints can indeed enlarge the IA feasible
region and enhance the system performance in both DoF and throughput
sense. Moreover, note that for a $12\times 12$, $2$ stream per Tx-Rx
pair and fully connected interference network, at most total network
$22$ DoF can be achieved, the performance of the proposed scheme (50
DoF) show that partial connectivity can indeed be exploit to
significantly increase network total DoF.

\subsection{Performance w.r.t. Partial Connectivity Factors}
\begin{figure} \centering
\includegraphics[scale=0.5]{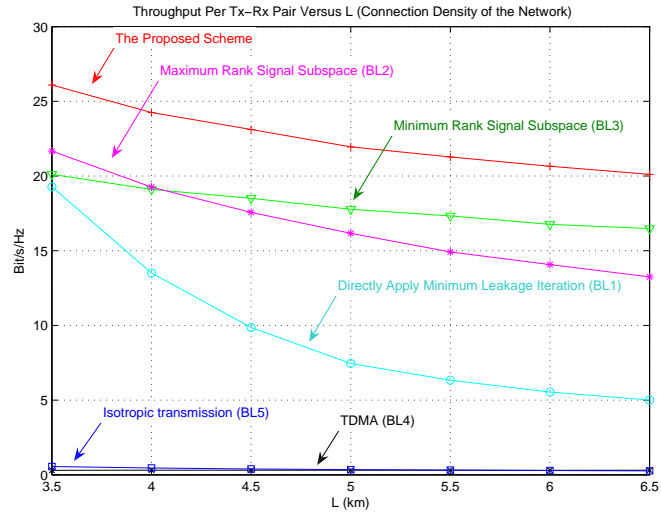}
\caption {Throughput per Tx-Rx  pair versus L(The maximum distance
that a Tx can interfere) for the proposed algorithm (and the 5
baselines) in a randomized partially connected MIMO interference
channel. The transmit power is given by $10log_{10}(P)=40 dB$ and
the local scattering radius $S=2.5km$.} \label{fig_perf_L}
\end{figure}

\begin{figure} \centering
\includegraphics[scale=0.5]{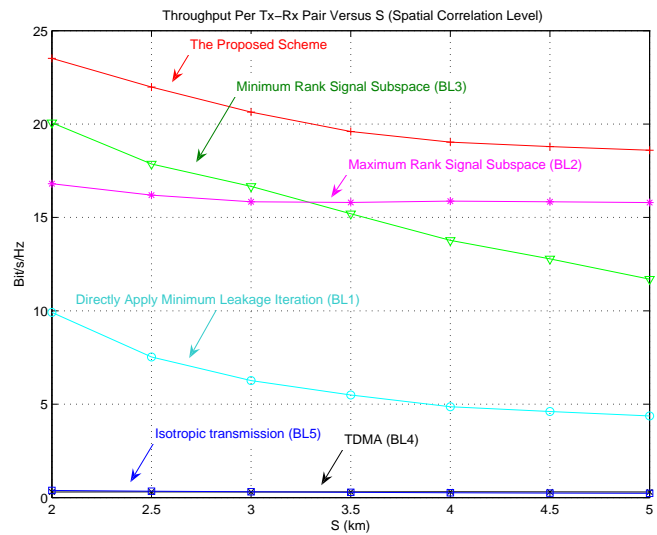}
\caption {Throughput per Tx-Rx  pair versus S(Radius of local
scattering) for the proposed algorithm (and the 5 baselines) in a
randomized partially connected MIMO interference channel. The
transmit power is given by $10log_{10}(P)=40 dB$ and $L=5km$.}
\label{fig_perf_S}
\end{figure}

To better illustrate how different partial connectivity factors such
as path loss and spatial correlation affect system performance, we
illustrate the sum throughput versus $L$ (the maximum distance that
a Tx can interfere a Rx) and $S$ (the radius of the local
scattering) under a fixed SNR (40dB) in Fig.~\ref{fig_perf_L} and
Fig.~\ref{fig_perf_S}, respectively. By comparing the performance of
the proposed scheme with different partial connectivity parameters,
we have that the performance of the proposed scheme scales
$\mathcal{O}\left(\frac{1}{LS}\right)$, which illustrates a
consistent observation as in Remark~\ref{remark:result} that weaker
partial connectivity can indeed contribute to higher system
performance. Moreover, comparison of the proposed algorithm with
Baseline 2 and 3 illustrates how we should select signal subspaces
$\mathbb{S}^{t*}_n$ and $\mathbb{S}^{r*}_m$ under different partial
connectivity regions. For example, low rank subspace is more
effective at high spatial correlation (small $S$) while high rank
subspace is more effective at low spatial correlation (large $S$).
Low rank subspace is also more effective compare to high rank
subspace in dense networks (large $L$) and { vice}
versa. By dynamically selecting signal subspace according to the
partial connectivity state of the network, the proposed scheme
obtains significant performance gain over a wide range of partial
connectivity levels.

\section{Conclusion}
In this paper, we have investigated how the partial connection can
be utilized to benefit the system performance in MIMO interference
networks. We considers a general partial connection model which
embraces various practical situations such as path loss effects and
spatial correlations. We proposed a novel two-stage interference mitigation
scheme. The stage I algorithm
 determines the stream assignment and the subspace
constraints for the precoders and decorrelators based on the partial connectivity state.
 The stage II algorithm
determines the precoders and decorrelators based on the stream
assignment and the subspace constraints as well as the local channel
state information. The signal spaces is designed to mitigate ``many"
IA constraints at a cost of only a ``few" free-variables in
precoders and decorrelator so as to extend the feasibility region of
the IA scheme. Analysis shows the proposed algorithm can
significantly increase system DoF in { symmetric}
partially connected MIMO interference networks. We also compare the
performance of  the proposed scheme with various baselines and show
via simulations that the proposed algorithms could achieve
significant gain in system performance of randomly connected
interference networks.

 \appendices
\section{Physical Interpretation for Virtual Angle Model in MIMO Channel}
\label{channel}

{ As has been observed in many previous works
\cite{AngleC1},\cite{AngleC3} the statistical property of the
channel states in a MIMO system is strongly affected by the physical
propagation environment. For instance, in cellular MIMO systems, the
Txs are positioned at high elevations above the scatterers while the
Rxs are
 positioned at low altitude with rich scattering
(Fig.~\ref{fig_channel1}A).
  Hence, only the scattering objects surrounding
a Rx could effectively reflect signals from the Tx to the Rx as
illustrated in Fig.~\ref{fig_channel1}B.} Following an approach
similar to \cite{Local2,Local3}, we assume the double directional
channel response from the $n$-th Tx to the $m$-th Rx
($n,m\in\{1,2,...,K\}$):
$\mathbf{H}_{mn}^a=\{h_{mn}^a(\theta_t,\theta_r),\;\theta_t,\theta_r\in[-{\pi,\pi})\}$
has the following property:
\begin{eqnarray}
h_{mn}^a(\theta_t,\theta_r)&=& 0 \mbox{ if } |\theta_t|>
\frac{\alpha}{2}; \label{eqn:double_1}
\\\mbox{where }\alpha&=&\left\{\begin{array}{l}
2\arcsin(\frac{S}{d_{mn}})\;\;\mbox{when: }S\le{d_{mn}}
\\2\pi\;\;\;\mbox{else.}\end{array}\right.\label{eqn:alpha_1}
\end{eqnarray}
where $d_{mn}$ is the distance between Tx {$n$} and to
{ Rx {$m$}}, $S$ is the \emph{local
effective scattering radius}. Assume the Tx are equipped with
uniform linear antenna array (ULA). Hence the {\em virtual angular
channel representation} \cite{AngleC2} is given by:
\begin{eqnarray}
h^v_{mn}(p,q)=\int_{-\pi}^{\pi}\int_{-\pi}^{\pi}
h_{mn}(\theta_t,\theta_r)f_{N_r}\left(\frac{\sin(\theta_r-\varphi_{mn})r}{\lambda}-\frac{p}{N_r}\right)
f_{N_t}\left(\frac{\sin(\theta_t-\theta_{mn})r}{\lambda}-\frac{q}{N_t}\right)
d\theta_rd\theta_t \label{eqn:vir_double}
\end{eqnarray}
where $r$ is the antenna separation,  $\lambda$ is the wavelength,
$\theta_{mn}$ is the angle between the transmit array normal
direction and the direction from Tx $n$ to Rx $m$, $\varphi_{mn}$ is
the angle between receive array normal direction and the direction
from Tx $n$ to Rx $m$, assume the antenna array is critically
spaced, i.e. $\frac{r}{\lambda}=\frac{1}{2}$.
$f_N(\omega)={\mathbf{e}^H_N(0)}\mathbf{e}_N(\omega)
=\frac{1}{N}e^{-j\pi\omega(N-1)}\frac{\sin(\pi N \omega)} {\sin(\pi
\omega)}$.

\begin{figure} \centering
\includegraphics[scale=0.4]{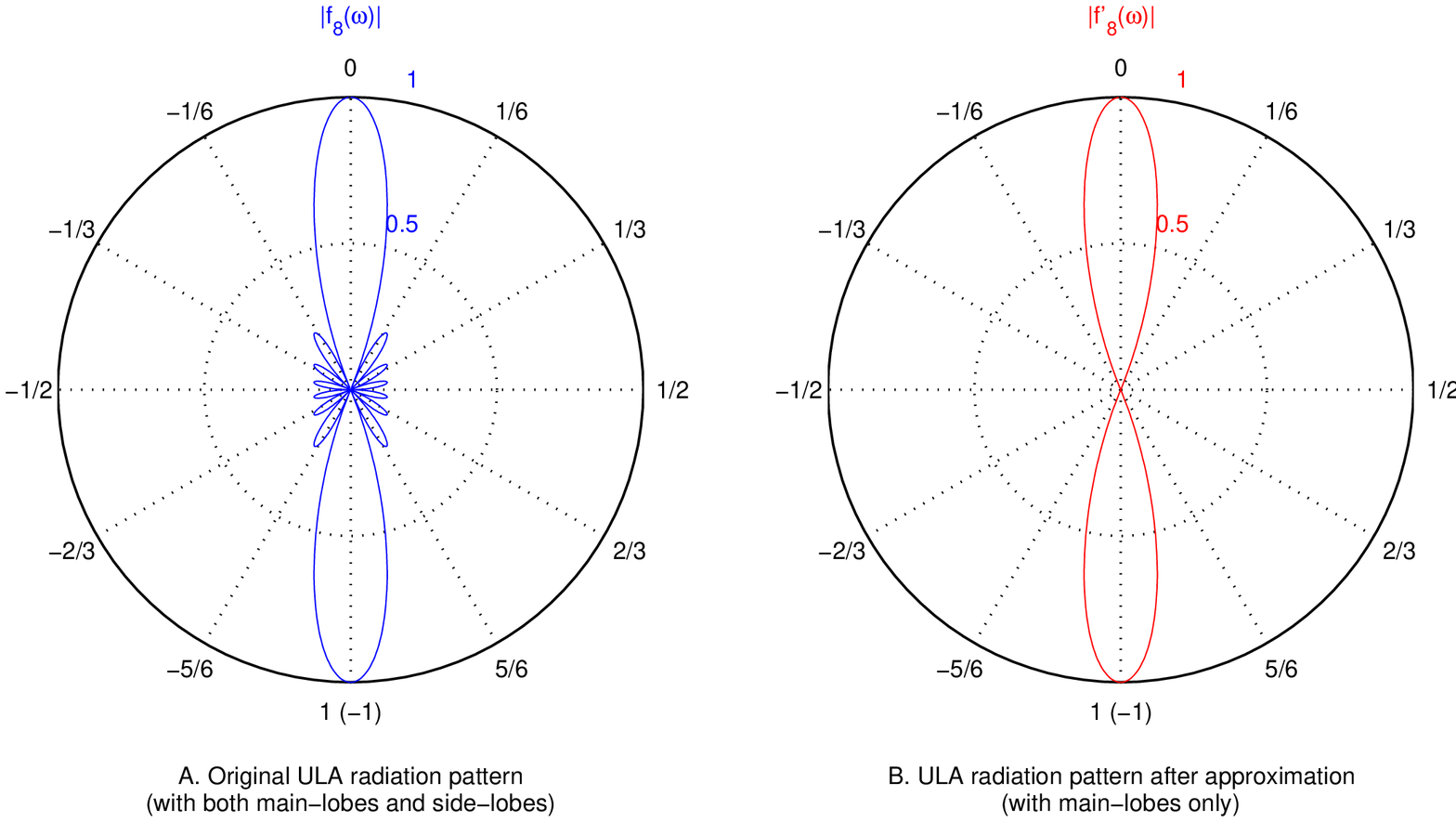}
\caption {Radiation pattern of ULA before and after approximation.}
\label{fig_polar}
\end{figure}

{ As illustrated in Fig.~\ref{fig_polar}A,
$f_N(\omega)$ represents the radiation pattern of ULA \cite{LA}.
Note that the main-lobes dominate in the radiation pattern (e.g.
when $N=8$, the power of the main-lobes occupy $91\%$ of that in the
whole radiation pattern). Hence, as illustrated in
Fig.~\ref{fig_polar}B, for simplicity, we use the main-lobe to
approximate the radiation pattern, i.e. we use
\begin{eqnarray}
f'_N(\omega)=\left\{\begin{array}{l}f_N(\omega), \mbox{ if: }
\omega-\lfloor\omega\rfloor\le \frac{1}{N} \mbox{ or
}\ge\frac{N-1}{N}\\0, \mbox{ otherwise.}\end{array}\right.
\end{eqnarray}
to replace $f_N(\omega)$ in \eqref{eqn:vir_double}.} Hence,
combining \eqref{eqn:double_1}, \eqref{eqn:alpha_1} and
\eqref{eqn:vir_double}, we have \eqref{eqn:virtual_1}.

\section{The Number of Freedoms and IA Constraints in Equations \eqref{eqn:cs_rank_1} and \eqref{eqn:cs_cross_1}}
\label{sec:proper} The freedoms in $\mathbf{V}'_{n}$ and
$\mathbf{U}'_{m}$ of \eqref{eqn:cs_rank_1} and
\eqref{eqn:cs_cross_1} are given by:
\begin{eqnarray}\mbox{dim}(\mathcal{G}(d_n,S^t_n))=d_n(S^t_n-d_f),\mbox{
and } \mbox{dim}(\mathcal{G}(d_m,S^r_m))=d_m(S^r_m-d_m), \mbox{
respectively.}\label{eqn:numv1}\end{eqnarray} where the Grassmannian
$\mathcal{G}(x,y)$ \cite{Grass1,Grass2} denotes the set of all
$x$-dimensional subspaces in $\mathbb{C}^{y}$.

Then consider the number of independent constraints in \eqref{eqn:cs_cross_1}.
Consider the singular value
decomposition of
$\mathbf{H}_{mn}=\mathbf{U}_{mn}\mbox{diag}(s_1,s_2,...,s_N)\mathbf{V}^H_{mn}$,
where $\mathbf{U}_{mn},\mathbf{V}_{mn}$ are $N_r\times N_r$ and
$N_t\times N_t$ unitary matrices, respectively, $N=\min(N_t,N_r)$,
$s_1\sim s_N$ are the singular values of $\mathbf{H}_{mn}$ in
descending order. Suppose $\mbox{rank}(\mathbf{H}_{mn})=r$, then we
have: $\mathbf{H}_{mn}=[\mathbf{U}^1_{mn},\mathbf{U}^2_{mn}] $ $
\left[\begin{array}{c@\;c}\mbox{diag}(s_1,s_2,...,s_r)&\mathbf{0}\\\mathbf{0}&\mathbf{0}\end{array}\right][
\mathbf{V}^1_{mn},\mathbf{V}^2_{mn}]^H$, where $\mathbf{U}^1_{mn}$
and $\mathbf{V}^1_{mn}$ are $N_r\times r$ and $N_t\times r$
matrices, respectively. Note that
\begin{eqnarray}
\mathbf{U}'_{m}\mathbf{S}^r_m\mathbf{H}_{mn}
\mathbf{S}^t_n\mathbf{V}'_{n}=\mathbf{0}\Leftrightarrow
\mathbf{U}'_{m}\mathbf{S}^r_m\mathbf{U}^1_{mn}\mbox{diag}(s_1,s_2,...,s_r)(\mathbf{V}^1_{mn})^H
\mathbf{S}^t_n\mathbf{V}'_{n}=\mathbf{0}
\end{eqnarray}
and
$\mbox{span}(\mathbf{V}^1_{mn})=(\mbox{span}(\mathbf{V}^2_{mn}))^\bot=(\mathcal{N}(\mathbf{H}_{mn}))^\bot$,
$\mbox{span}^H(\mathbf{U}^1_{mn})=(\mbox{span}^H(\mathbf{U}^2_{mn}))^\bot=(\mathcal{N}^H(\mathbf{H}_{mn}))^\bot$,
where $\mbox{span}(\mathbf{X})$, $\mbox{span}^H(\mathbf{X})$ denote
the linear space spanned by the columns of $\mathbf{X}$ and the rows
of $\mathbf{X^H}$, respectively. Hence, the number of independent
constraints in \eqref{eqn:cs_cross_1} is given by:
\begin{eqnarray}\mbox{rank}(\mathbf{U}'_{m}\mathbf{S}^r_m\mathbf{U}^1_{mn})
\cdot \mbox{rank}( \mathbf{V}^1_{mn}\mathbf{S}^t_n\mathbf{V}'_{n})=
\min\left(d_m,|\mathbb{S}^r_m\cap(\mathcal{N}^H(\mathbf{H}_{mn}))^\bot|\right)\min\left(
|\mathbb{S}^t_n\cap\mathcal{N}(\mathbf{H}_{mn})^\bot|, d_n
\right)\label{eqn:numc1}\end{eqnarray}

\section{Low Complexity Feasibility Checking Algorithm
in Step 4 of Stage~\ref{alg:offline}} \label{sec:alg_detail}
\begin{figure} \centering
\includegraphics[scale=0.78]{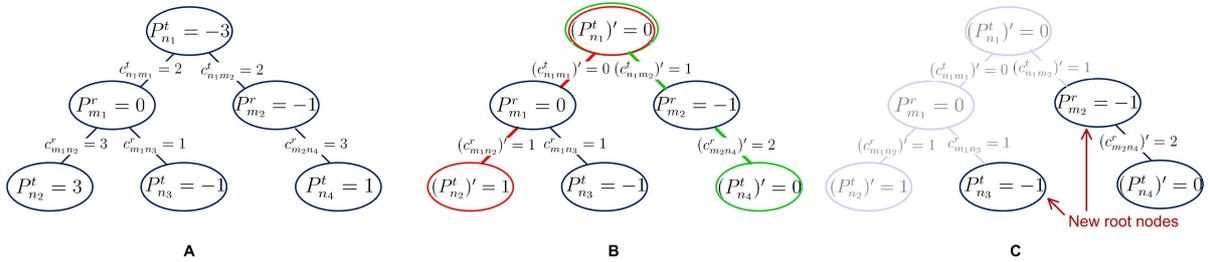}
\caption {Illustrative example of the ``pressure
transfer tree" and the corresponding operations in
Appendix~\ref{sec:alg_detail}. A) A tree generated in Step A and B;
B) Pressure transfer in Step C; C) Removal of depleted links and
neutrialized roots in Step D.} \label{fig_tree}
\end{figure}

{
\begin{itemize}
\item{\bf Initialize the constraint assignment:} Randomly generalize
a \emph{constraint assignment policy}, i.e. $\{c^t_{mn},c^r_{mn}\}$
such that: $c^t_{nm},c^r_{mn}\in\mathbb{N}\cup\{0\}$, $
c^t_{nm}+c^r_{mn}=c_{mn}$, $m,n\in\{1,2,...,K\}$ (Note that in the
subscripts of $\{c^t_{nm}\}$, transmitter indexes come first).
Calculate the \emph{variable - assigned constraint pressure}, i.e.
$\{P^t_{n},P^r_{m}\}$, where
\begin{eqnarray}P^t_{n}=v^t_{n}-\sum_{m\in\{1,2,...,K\}}c^t_{nm},\mbox{ and }
P^r_{m}=v^r_{m}-\sum_{n\in\{1,2,...,K\}}c^r_{mn}.\label{eqn:pressure}\end{eqnarray}
\item{\bf Update the constraint assignment:} While there exist ``overloaded nodes", i.e. $P^t_{n}<0$ or
$P^r_{m}<0$, $m,n\in\{1,2,...,K\}$, do the following to update
constraint assignment $\{c^t_{mn},c^r_{mn}\}$:
\begin{itemize}\item{\bf A. Initialization:}
Select an ``overloaded node" with negative pressure, without losing
generality, assume this node is Tx-{n}, $P^t_{n}<0$. Set $P^t_{n}$
to be the root node of the ``pressure transfer tree", which is
variation of the tree data structure, with its nodes storing the
pressures at the Txs and Rxs, its link strengths storing the maximum
number of constraints that can be reallocated between the parent
nodes and the child nodes. Please refer to Fig.~\ref{fig_tree} for
an example.
\item{\bf B. Add Leaf nodes to the
pressure transfer tree:}

For every leaf nodes (i.e. nodes without child nodes) $P^x_{n}$
($x\in\{t,r\}$, $n\in\{1,2,...,K\}$) with depths equal to the height
of the tree (i.e. the nodes at the bottom in Fig.~\ref{fig_tree}):
\begin{itemize}
\item[] For every $m\in\{1,2,...,K\}$: If $c^{\overline{x}}_{nk}>0$,
add $P^{\overline{x}}_{m}$ as a child node of $P^x_{n}$ with link
strength $c^{\overline{x}}_{nk}$, where $\overline{x}$ is the
element in $\{t,r\}$ other than $x$.
\end{itemize}
\item{\bf C. Transfer pressure from root to leaf nodes:} For every leaf node just
added to the tree in Step B with positive pressure, transfer
pressure from root to these leafs by updating the constraint
assignment policy $\{c^t_{mn},c^r_{mn}\}$. For instance, as
illustrated in Fig.~\ref{fig_tree}B,
$P^t_{n_1}\xrightarrow{c^t_{n_1m_1}}P^r_{m_1}\xrightarrow{c^r_{m_1n_2}}P^t_{n_2}$
is a root-to-leaf branch of the tree (red lines). Transfer pressure
from $P^t_{n_1}$ to $P^t_{n_2}$ by updating: $(c^t_{n_1m_1})'=
c^t_{n_1m_1}-\epsilon$, $(c^{r}_{m_1n_1})'=
c^{r}_{m_1n_1}+\epsilon$, $(c^r_{m_1n_2})'= c^r_{m_1n_2}-\epsilon$,
$(c^{t}_{n_2m_1})'= c^{t}_{n_2m_1}+\epsilon$. Hence we have
$(P^t_{n_1})'=P^t_{n_1}-\epsilon$ and
$(P^{t}_{n_2})'=P^{t}_{n_2}+\epsilon$, where $\epsilon$ is the
minimum of the absolute value of the root pressure, leaf pressure,
and all the strengths of the links, i.e. $\epsilon =
\min\left(-P^t_{n_1}, P^t_{n_2}, c^t_{n_1m_1}, c^r_{m_1n_2}\right)$,
$A'$ denotes the value of $A$ after update. Similarly, this
operation can also be done for the green lines in
Fig.~\ref{fig_tree}B.
\item{\bf D. Remove the ``depleted" links and ``neutralized" roots:}
\begin{itemize}
\item If the strength of a link become 0 after Step C: Separate the
subtree rooted from the child node of this link from the original
pressure transfer tree.
\item If the root of a pressure transfer tree (including the
subtrees just separated from the original tree) is nonnegative,
remove the root and hence the subtrees rooted from each child node
of the root become new trees. Repeat this process until all roots
are negative. For each newly generated pressure transfer tree,
repeat Steps B$\sim$D (Please refer to Fig.~\ref{fig_tree}C for an
example).
\end{itemize}
\item{\bf E. Exit Conditions:} Repeat Steps A$\sim$D until all
trees become empty (hence the network is IA feasible) or no new leaf
node can added for any of the non-empty trees in Step B (hence the
network is IA infeasible). Exit the algorithm.
\end{itemize}
\end{itemize}
}

\section{Proof for Lemma~\ref{thm:proper}}
{ \label{pf_thm:proper} We shall first prove the ``if"
side. From Step 4 in Stage~\ref{alg:offline} and the Initialization
step in Appendix~\ref{sec:alg_detail}, \eqref{eqn:proper_1} can be
rewritten as:
\begin{eqnarray}
\sum_{(n,m)\in\mathbb{G}}(c^t_{nm}+c^r_{mn})=\sum_{(n,m)\in\mathbb{G}}c_{mn}\le
\sum_{n\in\mathbb{G}_n}v^t_n+\sum_{m\in\mathbb{G}_m}v^r_m\label{eqn:proper_2}
\end{eqnarray}
$\forall \mathbb{G}_n,\mathbb{G}_m\subseteq \{1,2,...,K\}$,
$\mathbb{G}= \mathbb{G}_m\oplus\mathbb{G}_n$. From the exit
condition (Step E) of low complexity IA feasibility checking
algorithm, we have $P^t_{n}\ge0$, $P^r_{m}\ge0$, $\forall n,n\in
\{1,2,...,K\}$. Hence we have:
\begin{eqnarray}
\sum_{n\in\mathbb{G}_n}v^t_n+\sum_{m\in\mathbb{G}_m}v^r_m-
\sum_{(n,m)\in\mathbb{G}}(c^t_{nm}+c^r_{mn})&=&
\sum_{n\in\mathbb{G}_n}(v^t_n-\sum_{m\in\mathbb{G}_m}c^t_{nm})+
\sum_{m\in\mathbb{G}_m}(v^r_m-\sum_{n\in\mathbb{G}_n}c^r_{mn})\nonumber
\\&\ge&\sum_{n\in\mathbb{G}_n}(v^t_n-\sum_{m\in\{1,\atop2,...,K\}}c^t_{nm})+
\sum_{m\in\mathbb{G}_m}(v^r_m-\sum_{n\in\{1,\atop
2,...,K\}}c^r_{mn})\nonumber
\\&=&\sum_{n\in\mathbb{G}_n}P^t_n +\sum_{m\in\mathbb{G}_m}P^r_m\ge0
\label{eqn:ifside}
\end{eqnarray}
$\forall \mathbb{G}_n,\mathbb{G}_m\subseteq \{1,2,...,K\}$,
$\mathbb{G}= \mathbb{G}_m\oplus\mathbb{G}_n$. This completes the
``if" side proof.

Then we turn to the ``only if" side. We shall try to prove the
converse-negative proposition of the original statement. If the
network cannot pass the low complexity IA feasibility test, from the
exit condition (Step E), there must exists a non-empty pressure
transfer tree such that:
\begin{itemize}
\item Root node has negative pressure.
\item All other nodes are non-positive. This is because positive nodes are either
``neutralized" by the root in Step C if the strength of the links
from the root to these nodes are sufficient or separated from the
tree in Step D if one of the link strength is not sufficient.
\item No other nodes can be added to the tree, which implies $c^r_{mn}=0$ and $c^t_{m'n'}=0$ for any
Tx-n, Rx-m'  in the tree and Rx-m, Tx-n' not in the tree.
\end{itemize}
Hence, set $\mathbb{G}_n,\mathbb{G}_m$ in \eqref{eqn:proper_2} to be
the indexes of the Txs and Rxs that are in the remaining pressure
transfer tree, we have:
\begin{eqnarray}
&&\sum_{n\in\mathbb{G}_n}(v^t_n-\sum_{m\in\mathbb{G}_m}c^t_{nm})+
\sum_{m\in\mathbb{G}_m}(v^r_m-\sum_{n\in\mathbb{G}_n}c^r_{mn})=
\sum_{n\in\mathbb{G}_n}(v^t_n-\sum_{m\in\{1,2,...,K\}}c^t_{nm})+
\nonumber
\\&&
\sum_{m\in\mathbb{G}_m}(v^r_m-\sum_{n\in\{1,2,...,K\}}c^r_{mn})=\sum_{n\in\mathbb{G}_n}P^t_n
+\sum_{m\in\mathbb{G}_m}P^r_m<0 \label{eqn:otherside}
\end{eqnarray}
Hence, the network does not satisfy \eqref{eqn:proper_2}. This
completes the ``only if" side proof.

Finally, let us consider the complexity of the checking algorithm.
Since there are only $2K$ nodes, the algorithm can at most generate
$2K$ trees. For each tree, since each cross link can be added into
the tree once, there are at most $K(K-1)$ times of adding node
operation. Hence the worst case complexity is $\mathcal{O}(K^3)$.}

\section{Detail Modeling of
Symmetric Partially Connected MIMO Interference Network in
Definition~\ref{def:model}  } \label{sec:reason}
\begin{itemize}
\item {\bf Partial Connectivity due to Path Loss:}
If {$L<|n-m|<K-L$}  assume $\mathbf{H}_{mn} =
\mathbf{0}$.
\item {\bf Partial Connectivity due to Local Scattering (Direct Link):}
If $n=m$, assume due to local scattering,
$\mathbf{h}_{mn}^{ang}(p)=\mathbf{0}$ if: $ p \not\in \mathbb{E}_1$,
otherwise $\mathbf{h}_{mn}^{ang}(p)\sim \mathcal{CN}^{1\times N_t}$,
where $\mathbf{h}_{mn}^{ang}(p)$ is the $p$-th column of the angular
representation of the channel state $\mathbf{H}^{ang}_{mn}$ (defined
in \eqref{eqn:ant_vir}), $p\in\{0,1,...,N_t-1\}$,
$\mathbb{E}_1\subseteq\{0,1,...,N_t-1\}$ are the indices of the
{``good"}  angles. Denote $|\mathbb{E}_1|=E_1$.
\item {\bf Partial Connectivity due to Local Scattering (Cross Link):}
If {$0<|n-m|\le L$ or $|n-m|\ge K-L$} and $n\neq m$,
assume due to {local scattering}:
$\mathbf{h}_{mn}^{ang}(p)=\mathbf{0}$ if: $p \not\in
\mathbb{E}_2^{\triangle n}$, otherwise $\mathbf{h}_{mn}^{ang}(p)\sim
\mathcal{CN}^{1\times N_t}$, where $\triangle n=
n-m\in\{-L,-L+1,...L\}$, $\mathbb{E}_2^{\triangle n}$ are random
subsets of  $\{0,1,...,N_t-1\}$ which satisfy
$|\mathbb{E}_2^{\triangle n}|=E_2$, $\forall \triangle n$. $0\le
E_2\le N_t$.
\end{itemize}

\section{Proof for Theorem~\ref{thm:per1}}
\label{pf_thm:per1} Due to the symmetry property of the system, when
$\mathbb{D}^*=\{d_f,d_f,...,d_f\}$:
$\mathbb{S}^{t*}_n=\mathbb{S}^{t*}_m$,
$\mathbb{S}^{r*}_n=\mathbb{S}^{r*}_m$, $\forall n,m\in\{1,2,...,K\}$
and hence the system satisfies \eqref{eqn:proper_1} if and only if:
\begin{eqnarray}
\nonumber &&\sum_{n=1}^{K}\sum_{m\neq
n}^K\min\left(|\mathbb{S}^{r*}_m\cap(\mathcal{N}^H(\mathbf{H}_{mn}))^\bot|,d_f\right)
\min\left( |\mathbb{S}^{t*}_n\cap\mathcal{N}(\mathbf{H}_{mn})^\bot|,
d_f \right)
\\\nonumber &&\le
\sum_{n=1}^Kd_f(S^{t*}_n-d_f)+\sum_{m=1}^Kd_f(S^{r*}_m-d_f)
\\\nonumber &\Leftrightarrow&
\sum_{m=2}^K\min\left(S^{r*}-|\mathbb{S}^{r*}\cap\mathcal{N}^H(\mathbf{H}_{m1})|,d_f\right)\min\left(
S^{t*}-|\mathbb{S}^{t*}\cap\mathcal{N}(\mathbf{H}_{m1})|, d_f
\right)
\\&& \le d_f(S^{t*}+S^{r*}-2d_f) \label{eqn:proper2}
\end{eqnarray}
where $\mathbb{S}^{t*}=\mathbb{S}^{t*}_n$,
$\mathbb{S}^{r*}=\mathbb{S}^{r*}_n$, $\forall n\in\{1,2,...,K\}$,
${S}^{t*}=|\mathbb{S}^{t*}|$, ${S}^{r*}=|\mathbb{S}^{r*}|$.

In general, due to the randomness in
$\mathcal{N}^H(\mathbf{H}_{m1})$ and $\mathcal{N}(\mathbf{H}_{m1})$
it is hard to obtain optimal ${S}^{t*}$ and ${S}^{r*}$. To obtain a
fundamental insight, we shall consider two extreme policies:
${S}^{t*}=d_f,{S}^{r*}=\min(N_r,E_1)$ (smallest subspace dimension
on the Tx side and largest subspace dimension on the Rx side) and
$S^{t*}=E_1,{S}^{r*}=\min(N_r,E_1)$ (largest subspace dimension on
both the Tx and the Rx side).
 On the receiver side, when ${S}^{r*}=\min(N_r,E_1)$,
 $\mathbb{S}^{r*}=(\mathcal{N}^H(\mathbf{H}_{11}))^\perp$. On the
 transmitter side, from \eqref{eqn:NT_a}, the $\mathbb{S}^t_n(d)$ obtained in Step 3A,
Algorithm~\ref{alg:offline}  shall have the following form:
$\mathbb{S}^t_n(d)=\mbox{span}\left(\mathbf{e}_{N_t}(\frac{p_1}{N_t}),\mathbf{e}_{N_t}(\frac{p_2}{N_t}),...
\mathbf{e}_{N_t}(\frac{p_d}{N_t})\right)$, where $p_d$ is the p-th
index in $\mathbb{P}=\{0,1,...,N_t-1\}$, in which the elements are
ordered w.r.t to the metric
$\sum_{m=2}^{K}\mathbf{1}(\mathbf{e}_{N_t}(\frac{p}{N_t})\in\mathcal{N}(\mathbf{H}_{1m}))$
in descending order. When $S^{t*}=d_f$, \eqref{eqn:proper2} become:
\begin{eqnarray}
\nonumber &&\sum_{m=2}^{K}
d_f-|\mathbb{S}^{r*}\cap\mathcal{N}^H(\mathbf{H}_{m1})| \le
\min(N_r,E_1)-d_f
\\\nonumber &\Leftrightarrow& d_f(K-1)-\sum_{d=1}^{d_f}\sum_{m=2}^{K}
\mathbf{1}(\mathbf{e}_{N_t}(\frac{p_d}{N_t})\in\mathcal{N}(\mathbf{H}_{1m}))
\le \min(N_r,E_1)-d_f
\\\nonumber &\Leftarrow&d_f(K-1)-\frac{d_f}{N_t}\sum_{d=1}^{N_t}\sum_{m=2}^{K}
\mathbf{1}(\mathbf{e}_{N_t}(\frac{p_d}{N_t})\in\mathcal{N}(\mathbf{H}_{1m}))
\le \min(N_r,E_1)-d_f
\\&\Leftarrow&
d_f\left(\min(K-1,2L)\frac{E_2}{N_t}\right) \le \min(E_1,N_r)-d_f
\label{eqn:proper3}
\end{eqnarray}

When $d^*=E_1$, \eqref{eqn:proper2} is simplified to:
\begin{eqnarray}
\nonumber && \min(K-1,2L)\cdot\min(E_2,d_f) \le E_1+\min(E_1,N_r)-2d_f
\\&\Leftarrow&\min(K-1,2L)\cdot d_f \le E_1+\min(E_1,N_r)-2d_f
\label{eqn:proper4}
\end{eqnarray}

From \eqref{eqn:proper3} and \eqref{eqn:proper4},
\eqref{eqn:result1} is obtained, which completes the proof.

\end{document}